**Modeling Defect-Level Switching for Nonlinear and Hysteretic Electronic Devices**

Jiahao Dong and R. Jaramillo*

Department of Materials Science and Engineering, Massachusetts Institute of Technology, Cambridge, MA 02139, USA

* Corresponding author: rjaramil@mit.edu

*Abstract*

Previously, we demonstrated hysteretic and persistent changes of resistivity in two-terminal electronic devices based on charge trapping and detrapping at immobile metastable defects [H. Yin, A. Kumar, J.M. LeBeau, and R. Jaramillo, Phys. Rev. Applied **15**, 014014 (2021)]; we termed these devices as defect-level switching (DLS) devices. DLS devices feature all-electronic resistive switching and thus are volatile because of the "voltage-time" dilemma. However, the dynamics of volatile resistive switches may be valuable for emerging applications such as selectors in crosspoint memory, and neuromorphic computing concepts. To design memory and computing circuits using these volatile resistive switches, accurate modeling is essential. In this work we develop an accurate and analytical model to describe the switching physics in DLS devices, based on the established theories of point defect metastability in $Cu(In,Ga)Se_2$ (CIGS) and II-VI semiconductors. The analytical nature of our model allows for time-efficient simulations of dynamic behavior of DLS devices. We model the time durations of SET and RESET programming pulses, which can be exponentially shortened with respect to the pulse amplitude. We also demonstrate the concept of inverse design: given desired resistance states, the width and amplitude of the programming signal can be chosen accordingly.

## Introduction

The trapping and detrapping of electronic charge at immobile point defects can change the conductivity of metal/insulator/metal (MIM) stacks. This was analyzed by Simmons and Verderber in 1967,[1] and their interpretation has been applied in many MIM material systems.[2–5] This class of devices are known as all-electronic resistive switches. We recently proposed and demonstrated a mechanism for all-electronic resistive switching based on capture and release of electronic charge at metastable point defects at $MoO_3$/CdS heterojunctions.[6] These metastable defects can exist in two distinct lattice configurations: a shallow- and a deep-double-donor state, and they will switch between these two configurations upon stimulation with light or voltage bias; we termed this effect as defect level switching (DLS). In the case of $MoO_3$/CdS heterojunctions, the DLS-active defects are the sulfur vacancies $V_S$ in CdS.[7] Similar DLS phenomenology has been observed in other compound semiconductors, including DX centers in III-Vs,[8–10] and metastable defect complexes in the Cu-In-Ga-S-Se system (CIGS).[11–14] In DLS devices, the state retention time is determined by point defect metastability; compared this to MIM stacks, in which the designed potential wells determine the retention time.

Because of the "voltage-time" dilemma analyzed by Schroeder *et al.*, all-electronic resistive switching is not suitable for nonvolatile memory: the high current densities required for short READ/WRITE pulse (~ns) will result in the retention time many orders of magnitude less than the benchmark value (~years).[15] However, for emerging applications such as selectors in a crosspoint array,[16] and neuromorphic computing concepts that operate in the temporal domain, such as reservoir computing,[17] the dynamical behavior of volatile resistive random access memory (RRAM) is valuable. For instance, Wang *et al.* utilized the dynamics of silver nanoparticles to functionally resemble the synaptic $Ca^{2+}$ behavior.[18] Using the volatile switching effect of silver filaments, Zhang *et al.* achieved the emulation of multiple essential features of artificial neuron in a unified manner without auxiliary circuits.[19] Quantitative modeling of the dynamics of conductive filament-based RRAM devices has been extensively studied, far more so than that for all-electronic RRAM.[20–25] All-electronic volatile RRAM may prove particularly useful if the switching dynamics can be predictively modelled based on established semiconductor physics concepts, and if devices not relying on mass transport prove more reliable in operation than devices relying on ion motion, nanoscale redox chemistry, and other complex coupled electronic-ionic phenomena. In this context, DLS devices occupy a unique niche: they feature dynamics that can be quantitatively modeled and tuned through materials selection and device design, and they operate on the basis of charge state transitions at immobile defects.

Here we develop an analytical and highly efficient model to describe the dynamics of DLS devices, starting from the statistics of point defect charge state transition. In **§1**, we describe the statistics of DLS-active point defects, following established theories of metastable point defects in CIGS solar cells,[11–14] and II-VI semiconductors.[7,26,27] In **§2**, we model a metal-semiconductor heterojunction with DLS-active defects, as we have demonstrated experimentally.[6] The resistive switching is caused by DLS defects that remain in metastable charge states after voltage stimulus, resulting in an out-of-equilibrium band diagram. Using Kimerling's model of deep traps, we derive an analytical expression of the interface band diagram by approximating the space charge profile using a step function.[28] The effects of bias treatment on the junction conductivity are consistent with our experimental results, and the heterojunction band diagram and metastable defect distributions computed from our model are in great agreement with those computed from numerical iteration (as proposed by Decock *et al.*).[14] Also, we estimate that our model is at least one million times more efficient compared to the numerical approach.

The analytical nature of our model allows for simulations of the device dynamics in a time-efficient manner. In **§3**, we illustrate the nonlinear and hysteretic resistive switching of a DLS heterojunction device upon a sinusoidal driving voltage. We model and discuss the hysteretic behavior at different frequencies of the driving voltage. We model the time durations of SET and RESET programming pulses, which can be exponentially shortened with respect to the pulse amplitude. These simulations demonstrate the concept of inverse design: given desired resistance states, the time width and voltage amplitude of the programming signal can be chosen accordingly. The concept of inverse design may extend to materials selection led by circuit design and system optimization, provided that a larger and accurate database of DLS-active defects becomes available.

## 1. Configuration transition kinetics of DLS-active defects

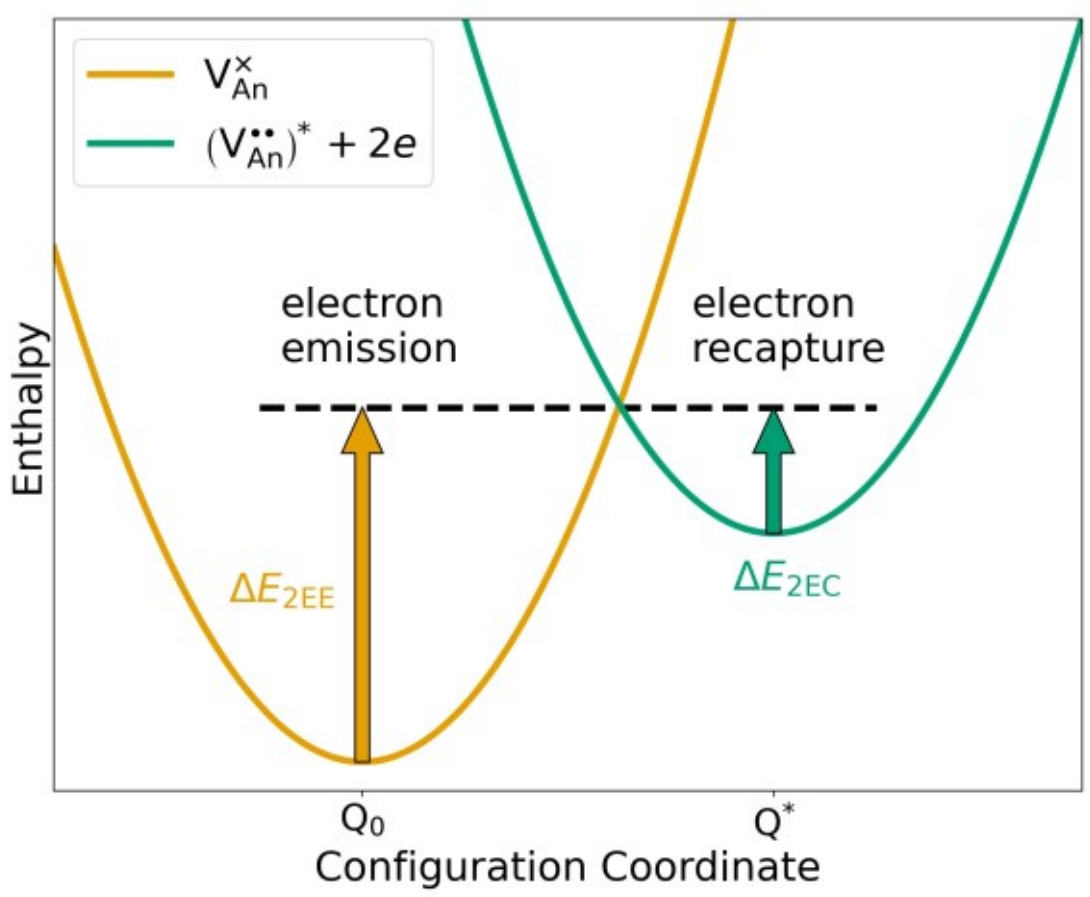

**Figure 1:** Configuration coordinate diagram for DLS-active point defects that exhibit metastable behavior – here, anion vacancies that are double donors. $Q_0$ and $Q^*$ represent stable configuration coordinates for the neutral deep donor state $\mathrm{V}_{\mathrm{An}}^{\times}$ and doubly-ionized shallow donor state $(\mathrm{V}_{\mathrm{An}}^{\bullet\bullet})^*$, respectively; the asterisk indicates a metastable lattice configuration. System enthalpy *vs.* configuration coordinate $\mathrm{V}_{\mathrm{An}}^{\times}$ and $(\mathrm{V}_{\mathrm{An}}^{\bullet\bullet})^*$ are represented by orange and green solid lines, respectively. Thermal activation energies $\Delta E_{2\mathrm{EE}}$ and $\Delta E_{2\mathrm{EC}}$ necessary for transitions in Eq. (1) and Eq. (4) are indicated by orange and green arrows, respectively.

DLS-active defects are found in a number of compound semiconductors, including DX centers in *n*-AlGaAs and III-nitrides,[8–10] the oxygen vacancy $\mathrm{V}_{\mathrm{O}}$ in *n*-ZnO,[26,27] the sulfur vacancy $\mathrm{V}_{\mathrm{S}}$ in *n*-CdS.[7] Though in this section we use specific features of anion vacancies in wide-bandgap II-VI semiconductors to discuss configuration transition kinetics, the model can be generalized to any other semiconductor system containing DLS-active defects.

In **Fig. 1** we present the configuration coordinate diagram for DLS-active defects, using for illustrative example the case of anion vacancies that are double donors. We represent the deep, neutral state of the anion vacancy as $\mathrm{V}_{\mathrm{An}}^{\times}$, which is occupied by two electrons, corresponding to $\mathrm{V}_{\mathrm{O}}^{\times}$ in ZnO and $\mathrm{V}_{\mathrm{S}}^{\times}$ in CdS; upon ionization, $\mathrm{V}_{\mathrm{An}}^{\times}$ can transform into the doubly-ionized, shallow donor state $(\mathrm{V}_{\mathrm{An}}^{\bullet\bullet})^*$; the asterisk indicates a metastable lattice configuration. In ZnO, due to the negative-*U* effects the singly-ionized state $\mathrm{V}_{\mathrm{O}}^{\bullet}$ is never thermodynamically stable.[26,29] In **§A1** we plot the equilibrium distributions of different charge states for the oxygen vacancy, and the concentration of singly-ionized state is at least ten orders of magnitude smaller than $\mathrm{V}_{\mathrm{O}}^{\times}$ or $(\mathrm{V}_{\mathrm{O}}^{\bullet\bullet})^*$. Therefore, the transitions $\mathrm{V}_{\mathrm{An}}^{\times}/\mathrm{V}_{\mathrm{An}}^{\bullet}$ and $\mathrm{V}_{\mathrm{An}}^{\bullet}/(\mathrm{V}_{\mathrm{An}}^{\bullet\bullet})^*$ are experimentally inaccessible, and we only address the configuration conversion between the neutral and doubly-ionized metastable states $\mathrm{V}_{\mathrm{An}}^{\times}$ and $(\mathrm{V}_{\mathrm{An}}^{\bullet\bullet})^*$.

In *n*-type semiconductors, the transition from deep to shallow donor configuration can happen via electron emission process, and the reaction path is given by:

$$\mathrm{V}_{\mathrm{An}}^{\times} \rightarrow (\mathrm{V}_{\mathrm{An}}^{\bullet\bullet})^* + 2e \tag{1}$$

The transition in Eq. (1) requires simultaneous emission of two electrons as well as thermal activation over an energy barrier $\Delta E_{2\mathrm{EE}}$, as indicated in the configuration coordinate diagram in

**Fig. 1**. Accompanying the deep-to-shallow configuration transition, the cation-cation distances around the anion vacancy may increase by upwards of 30%.[30] An energy of $\Delta E_{2\mathrm{EE}}$ is required to distort the lattice configuration, and shift the defect level to be resonant with the conduction band. The optical transition energy is higher than the thermal energy barrier, because crystal-momentum-conserving optical transitions can occur only at fixed configuration coordinate.[8] In addition to the reaction path described in Eq. (1), when valence band holes are injected through bias treatment in a junction, the deep-to-shallow transition is possible via simultaneous capture of one hole and emission of one electron, or simultaneous capture of two holes, with thermal energy barriers $\Delta E_{1\mathrm{HC}}$, and $\Delta E_{2\mathrm{HC}}$, respectively, and the corresponding reaction pathways are given by:

$$\mathrm{V}_{\mathrm{An}}^{\times} + h \rightarrow (\mathrm{V}_{\mathrm{An}}^{\bullet\bullet})^{*} + e \tag{2}$$

$$\mathrm{V}_{\mathrm{An}}^{\times} + 2h \rightarrow (\mathrm{V}_{\mathrm{An}}^{\bullet\bullet})^{*} \tag{3}$$

As shown in **Fig. 1**, the reverse transition from the metastable state to the neutral state happens through simultaneous capture of two electrons from the conduction band as well as thermal activation over an energy barrier $\Delta E_{2\mathrm{EC}}$, and the reaction path is given by:

$$(\mathrm{V}_{\mathrm{An}}^{\bullet\bullet})^{*} + 2e \rightarrow \mathrm{V}_{\mathrm{An}}^{\times} \tag{4}$$

In principle, the transition from the doubly-ionized state to the neutral state can happen through hole emission processes, *i.e.*, simultaneous capture of one electron and emission of one hole, or simultaneous emission of two holes. To emit holes to the valence band, the defect needs to be in acceptor states. Because the defects discussed in this work (anion vacancies in II-VI semiconductors) are either in shallow donor states or deep donor states, hole emission processes are unlikely to happen. However, in some other semiconductor systems, the contribution from hole emission needs to accounted for. For instance, the defect complexes in CIGS are amphoteric and can transform between donor and acceptor states.[13] In this case, the donor-to-acceptor configuration transition pathways should include those through hole emission processes.

The energy barriers related to these configuration transitions can be calculated by *ab initio* methods or inferred from experimental data. Using the local density approximation (LDA) of density function theory (DFT), Lany and Zunger calculated the configuration coordinate diagram for the oxygen vacancy in ZnO, which gives $\Delta E_{2\mathrm{EE}} = 3.7$ eV, $\Delta E_{2\mathrm{EC}} = 0.2$ eV and $\Delta E_{1\mathrm{HC}} = 1.3$ eV.[27] For $\mathrm{V_S}$ in CdS, using the experimental data on photoconductivity, we infer that the thermal activation energy for electron recombination is $\Delta E_{2\mathrm{EC}} = 0.6$ eV; using the experimental data on resistivity of $MoO_3$/CdS heterojunction devices, we infer $\Delta E_{2\mathrm{EE}} = 0.9$ eV.[6,7] In the presence of free holes, the deep donor DX state is immediately ionized without thermal activation, *i.e.*, $\Delta E_{2\mathrm{HC}} = 0$ for both defects.[31] An important future work for us is to use DFT calculations to accurately compute the configuration coordinate diagram for $\mathrm{V_S}$ as well as other DLS-active defects, and compare the calculations results to the experimentally determined activation energy barriers.

Now we discuss the kinetics of transitions described in Eqs. (1)-(4). According to the Shockley-Read-Hall (SRH) mechanism, we denote one-electron emission/capture rates as $\tau_{\mathrm{ee/ec}}^{-1}$, which are given by $\tau_{\mathrm{ee}}^{-1} = v_{\mathrm{th}} N_{\mathrm{C}} \sigma$ and $\tau_{\mathrm{ec}}^{-1} = v_{\mathrm{th}} n \sigma$, respectively, where $v_{\mathrm{th}}$ is the carrier thermal velocity, $N_{\mathrm{C}}$ the effective density of states of the conduction band, $n$ the electron density in the conduction band and $\sigma$ the carrier capture cross section. Since $n$ is given by $n = N_{\mathrm{C}}\exp\left(-(E_{\mathrm{C}} - E_{\mathrm{F}})/kT\right)$, where $E_{\mathrm{C}}$ is the energy level of the conduction band and $E_{\mathrm{F}}$ the Fermi level, we can rewrite the electron capture rate as $\tau_{\mathrm{ec}}^{-1} = \upsilon_{\mathrm{th}} N_{\mathrm{C}} \sigma \exp\left(-(E_{\mathrm{C}} - E_{\mathrm{F}})/kT\right)$. We denote the transition rates of Eqs. (1)-(4) as $\tau_{\mathrm{2EE}}^{-1}$, $\tau_{\mathrm{1HC}}^{-1}$, $\tau_{\mathrm{2HC}}^{-1}$ and $\tau_{\mathrm{2EC}}^{-1}$, respectively. For the transition in Eq. (1), the attempt frequency for activation over the energy barrier $\Delta E_{\mathrm{2EE}}$ is comparable to the phonon frequency $\upsilon_{\mathrm{ph}}$, and the probability of one successful activation attempt is $\exp(-\Delta E_{\mathrm{2EE}}/kT)$. Following Lany and Zunger's analysis, after one successful activation attempt, the time window for electron emission is $\upsilon_{\mathrm{ph}}^{-1}$, and during this time window the probability of two-electron emission is $\left(\upsilon_{\mathrm{ph}}^{-1}\tau_{\mathrm{ee}}^{-1}\right)^2 = \left(\upsilon_{\mathrm{ph}}^{-1}\upsilon_{\mathrm{th}} N_{\mathrm{C}} \sigma\right)^2$.[13] Hence, the transition rate $\tau_{\mathrm{2EE}}^{-1}$ is given by the following expression:

$$\tau_{\mathrm{2EE}}^{-1} = \frac{1}{\upsilon_{\mathrm{ph}}} (\upsilon_{\mathrm{th}} N_{\mathrm{C}} \sigma)^2 \exp\left(-\frac{\Delta E_{\mathrm{2EE}}}{kT}\right) \tag{5}$$

Here, the assumption is that the one-electron capture cross sections are the same in the first and second ionization processes. We acknowledge that it's a crude assumption, and it may lower the accuracy of simulation results. According to the analysis by Chicot *et al.*, the cross section is larger in the second ionization than the first one, based on an argument of the Huang and Rhys parameter.[32] However, more rigorous analysis and calculations are required to accurately determine the capture cross sections in the two ionization processes. As we will show in **§3**, Eq. (25) determines the device dynamics and the capture cross sections only enter the prefactor of Eq. (25), without affecting the hysteretic behavior and voltage dependence and of the device dynamics. Therefore, in this work we are not bothered unduly by this assumption.

Following our discussion on $\tau_{\mathrm{2EE}}^{-1}$, $\tau_{\mathrm{2EC}}^{-1}$, $\tau_{\mathrm{1HC}}^{-1}$ and $\tau_{\mathrm{2HC}}^{-1}$ are given by:

$$\tau_{\mathrm{2EC}}^{-1} = \frac{1}{\upsilon_{\mathrm{ph}}} (\upsilon_{\mathrm{th}} N_{\mathrm{C}} \sigma)^2 \exp\left(-\frac{2E_{\mathrm{C}} - 2E_{\mathrm{F}} + \Delta E_{\mathrm{2EC}}}{kT}\right) \tag{6}$$

$$\tau_{\mathrm{1HC}}^{-1} = \frac{1}{\upsilon_{\mathrm{ph}}} \left(\upsilon_{\mathrm{th}} \sqrt{N_{\mathrm{V}} N_{\mathrm{C}}} \sigma\right)^2 \exp\left(-\frac{E_{\mathrm{F}} - E_{\mathrm{V}} + \Delta E_{\mathrm{1HC}}}{kT}\right) \tag{7}$$

$$\tau_{\mathrm{2HC}}^{-1} = \frac{1}{\upsilon_{\mathrm{ph}}} (\upsilon_{\mathrm{th}} N_{\mathrm{V}} \sigma)^2 \exp\left(-\frac{2E_{\mathrm{F}} - 2E_{\mathrm{V}} + \Delta E_{\mathrm{2HC}}}{kT}\right) \tag{8}$$

where $N_{\mathrm{V}}$ is the effective density of states of the valence band.

We should note that in this work the charge transition rates are basically calculated by the SRH model, and we account for the lattice configuration changes by introducing Boltzmann factors. The SRH theory is a simplified model and it was shown to be invalid to describe the charge exchange rates between two materials.[33] In the latest works, advanced defect models such as the non-radiative multi-phonon (NMP) theory and the Marcus charge transfer theory combined with DFT calculations have been used to accurately calculate the charge transition rates.[33–36] Integration of these advanced defect models into our model will effectively improve the accuracy of simulation results, that we leave in our future work.

We denote the shallow-to-deep and deep-to-shallow configuration transition rates as $\tau_{\mathrm{S\to D}}^{-1}$ and $\tau_{\mathrm{D\to S}}^{-1}$, respectively. For the shallow-to-deep transition, Eq. (4) is the only pathway considered in this work. For the deep-to-shallow transition, the two-hole capture process in Eq. (3) is the dominant mechanism because it has the lowest thermal activation energy. The other processes have higher energy barriers and can gain importance at high temperatures. In **§A1** we analyze $\tau_{\mathrm{2HC}}^{-1}$, $\tau_{\mathrm{1HC}}^{-1}$ and $\tau_{\mathrm{2EE}}^{-1}$ as a function of Fermi energy and demonstrate that $\tau_{\mathrm{2HC}}^{-1}$ is the most important contribution upon junction reverse biasing and hole injection. Therefore, $\tau_{\mathrm{S\to D}}^{-1} = \tau_{\mathrm{2EC}}^{-1}$ and $\tau_{\mathrm{D\to S}}^{-1} = \tau_{\mathrm{2HC}}^{-1}$, which also preserves the symmetry of rate equations:

$$\tau_{\mathrm{S\to D}}^{-1} = \tau_{\mathrm{2EC}}^{-1} = \frac{1}{\upsilon_{\mathrm{ph}}}(V_{\mathrm{th}} N_{\mathrm{C}} \sigma)^2 \exp\left(-\frac{2E_{\mathrm{C}} - 2E_{\mathrm{F}} + \mathit{\Delta} E_{\mathrm{2EC}}}{kT}\right) \quad (9)$$

$$\tau_{\mathrm{D\to S}}^{-1} = \tau_{\mathrm{2HC}}^{-1} = \frac{1}{\upsilon_{\mathrm{ph}}}(V_{\mathrm{th}} N_{\mathrm{V}} \sigma)^2 \exp\left(-\frac{2E_{\mathrm{F}} - 2E_{\mathrm{V}} + \mathit{\Delta} E_{\mathrm{2HC}}}{kT}\right) \quad (10)$$

We denote the Fermi level position where the deep donor state $\mathrm{V}_{\mathrm{An}}^{\times}$ and the metastable shallow donor state $(\mathrm{V}_{\mathrm{An}}^{\bullet\bullet})^*$ are equally likely as the transition energy $E_{\mathrm{trans}}$, *i.e.*, $\tau_{\mathrm{D\to S}}^{-1} = \tau_{\mathrm{S\to D}}^{-1}$ when $E_{\mathrm{F}} = E_{\mathrm{trans}}$. Using Eq. (9) and Eq. (10), we have:

$$E_{\mathrm{trans}} = E_{\mathrm{C}} - \frac{E_{\mathrm{g}}}{2} + \frac{\Delta E_{\mathrm{2EC}} - \Delta E_{\mathrm{2HC}}}{4} \quad (11)$$

where $E_{\mathrm{g}}$ is the semiconductor band gap. For $E_{\mathrm{F}} > E_{\mathrm{trans}}$, $\tau_{\mathrm{D\to S}}^{-1} < \tau_{\mathrm{S\to D}}^{-1}$ and the deep donor state $\mathrm{V}_{\mathrm{An}}^{\times}$ is thermodynamically favored; for $E_{\mathrm{F}} < E_{\mathrm{trans}}$, $\tau_{\mathrm{D\to S}}^{-1} > \tau_{\mathrm{S\to D}}^{-1}$ and the metastable shallow donor state $(\mathrm{V}_{\mathrm{An}}^{\bullet\bullet})^*$ is thermodynamically favored.

If we denote the concentration of DLS-active defects as $N_{\mathrm{DLS}}$, and the concentration of deep and shallow donor states as $N_{\mathrm{deep}}$ and $N_{\mathrm{shallow}}$, respectively, the time variation of $N_{\mathrm{shallow}}(t)$ and $N_{\mathrm{deep}}(t)$ are given by:

$$\frac{\partial N_{\mathrm{shallow}}(t)}{\partial t} = \tau_{\mathrm{D\to S}}^{-1} N_{\mathrm{deep}}(t) - \tau_{\mathrm{S\to D}}^{-1} N_{\mathrm{shallow}}(t) \quad (12)$$

$$\frac{\partial N_{\mathrm{deep}}(t)}{\partial t} = \tau_{\mathrm{S\to D}}^{-1} N_{\mathrm{shallow}}(t) - \tau_{\mathrm{D\to S}}^{-1} N_{\mathrm{deep}}(t) \quad (13)$$

At steady states, the time variation of $N_{\mathrm{deep}}(t)$ and $N_{\mathrm{shallow}}(t)$ are both equal to zero, yielding steady state concentrations $N_{\mathrm{deep}}^{\mathrm{SS}}$ and $N_{\mathrm{shallow}}^{\mathrm{SS}}$:

$$N_{\mathrm{deep}}^{\mathrm{SS}} = \frac{1}{1 + \exp\left(\frac{2E_{\mathrm{trans}} - 2E_{\mathrm{F}}}{kT}\right)} N_{\mathrm{DLS}} \quad (14)$$

$$N_{\text{shallow}}^{\text{SS}} = \frac{1}{1 + \exp\left(-\frac{2E_{\text{trans}} - 2E_{\text{F}}}{kT}\right)} N_{\text{DLS}} \quad (15)$$

The equations in this section connect the dynamics of DLS devices to defect and material parameters, via fundamental semiconductor statistics. In **Table 1**, we list the properties of DLS-active defects in II-VI semiconductors, *i.e.*, $V_O^\times$ in ZnO and $V_S^\times$ in CdS, and other material parameters, which we will use in simulations in **§2** and **§3**. For future work, a computational database of DLS-active defects is required. The database should include defect parameters which are necessary to guide the design of DLS devices, such as thermal activation energies of configuration transitions. These properties could differ greatly from one defect to another, and should computed carefully and accurately by DFT calculations and calibrated by experiments. Such a computational database, combined with the model that we develop in this work, suggests promising opportunities for technology computer-aided design (TCAD) of resistive switching devices based on DLS phenomena.

| *Material* → *Parameter* ↓ | *CdS* | *ZnO* |
|---|---|---|
| $E_g$ (eV) | 2.4 | 3.4 |
| $\chi_s$ (V) | 4.3 | 4.5 |
| $\varepsilon_r$ | 9 | 11 |
| $N_C$ ($\text{cm}^{-3}$) | $2.4 \times 10^{18}$ | $2.7 \times 10^{18}$ |
| $V_{th}$ (cm/s) | $2.5 \times 10^7$ | $2.4 \times 10^7$ |
| $\mu_n$ ($\text{cm}^2/(\text{V} \cdot \text{s})$) | 2 | 200 |
| DLS-active defects | Sulfur vacancy $V_S$ | Oxygen vacancy $V_O$ |
| $E_C - E_{trans}$ (eV) | 1.1 | 1.7 |
| $\Delta E_{2EC}$ (eV) | 0.6 | 0.2 |
| $\Delta E_{2HC}$ (eV) | 0.0 | 0.0 |
| $\sigma$ ($\text{cm}^{-2}$) | $10^{-13}$ | $10^{-13}$ |
| $\upsilon_{ph}$ (THz) | 1 | 1 |
| $N_{DLS}$ ($\text{cm}^{-3}$) | $2 \times 10^{18}$ | |
| $N_d$ ($\text{cm}^{-3}$) | $10^{12}$ | |
| Thickness (um) | 0.06 | |

**Table 1:** Properties of two wide-band gap II-VI semiconductors that host DLS-active defects: $V_O^\times$ in ZnO and $V_S^\times$ in CdS. $E_C - E_{trans}$ is calculated according to Eq. (11). For the carrier capture cross section $\sigma$, Chicot *et al.* identified a deep donor of a large electron capture cross section of $1.6 \times 10^{-13}$ $\text{cm}^{-2}$ to be related to $V_O$ in ZnO.[32] The order of magnitude of $10^{-13}$ $\text{cm}^{-2}$ is in agreement with a positively charged defect center and we assign $\sigma = 10^{-13}$ $\text{cm}^{-2}$ for both $V_S$ in CdS and $V_O$ in ZnO. We take the phonon frequency $\upsilon_{ph} = 1$ THz for both defects. $N_{DLS}$ and $N_d$ are the concentrations of DLS-active defects and non-DLS-active shallow donors. We keep $N_{DLS}$ the same as the metastable defect concentration that we use in device simulations in our previous work.[6] The layer thickness is used in the simulations below.

## 2. Resistive switching physics in DLS-heterojunction devices

Configuration conversion between deep neutral and ionized states of DLS-active defects leads to persistent photoconductivity, and is useful in high-responsivity photoconductive sensors. However, it can also be achieved without illumination by applied bias in a metal-semiconductor Schottky junction, where the metal serves as a hole injection layer to induce the configuration transition defined in Eq. (3). Though operating at an interface, DLS devices are not based on the valence-change resistive switching mechanism in which crystal defects drift under an electric field, and there is no mass transport in DLS devices.[37] In case of the $MoO_3$/CdS heterojunction that we have experimentally demonstrated, $MoO_3$ has a higher work function than CdS and thus can inject holes into CdS, leading to a space charge layer (SCL) within CdS.[6] Within the SCL, most metastable defects are in the shallow donor state $(V_S^{\bullet\bullet})^*$. In **Fig. 2a** we present the energy levels of materials used to make a DLS-heterojunction device: an *n*-type semiconductor with DLS-active defects, and a metal or highly-doped semiconductor. $W_m$ and $q\chi_s$ refer to the metal work function and the semiconductor electron affinity, respectively. In the semiconductor, $E_{deep}$ and $E_{shallow}$ refer to the energy levels of deep and shallow donor states, respectively. Because of the negative-*U* effects, the first ionization energy $E_{deep}$ (corresponding to the $V_{An}^{\times}/V_{An}^{\bullet}$ transition) is deeper than $E_{trans}$ (corresponding to the $V_{An}^{\times}/(V_{An}^{\bullet\bullet})^*$ transition).[38] The Fermi level $E_F$ is above the transition energy $E_{trans}$. We denote by $q\phi_n$ the energy difference between $E_C$ and the Fermi level $E_F$. In **Fig. 2b** we present the schematic band diagram of an unbiased DLS-heterojunction. The formation of Schottky barrier $q\phi_S$ is due to the metal work function $W_m$ being larger than the semiconductor electron affinity $q\chi_s$. In the semiconductor side, the conduction band edge energy $E_C(x)$ and the metastable state transition energy $E_{trans}(x)$ vary with distance $x$ from the heterojunction. We denote by $x_{trans}$ the location where $E_{trans}(x)$ intersects the Fermi energy:

$$E_{trans}(x = x_{trans}) = E_F \tag{16}$$

In **Fig. 2b** the position of $x_{trans}$ is indicated by a black star. According to Eqs. (14)-(16), $x_{trans}$ is where both of the steady state concentrations $N_{deep}^{SS}(x)$ and $N_{shallow}^{SS}(x)$ are equal to $N_{DLS}/2$, *i.e.*, $V_{An}^{\times}$ and $(V_{An}^{\bullet\bullet})^*$ are equally likely at $x_{trans}$. Because the metastable state distribution $N_{shallow}^{SS}(x)$ affect the band bending and $E_{trans}(x)$, $N_{deep}^{SS}(x)$ and $N_{shallow}^{SS}(x)$ have to be calculated iteratively for self-consistency. The calculation procedure is as described by Decock *et al.*, in the context of studying apparent doping profiles in CIGS solar cells that contain metastable ($V_{Se}$-$V_{Cu}$) defect complexes.[14]

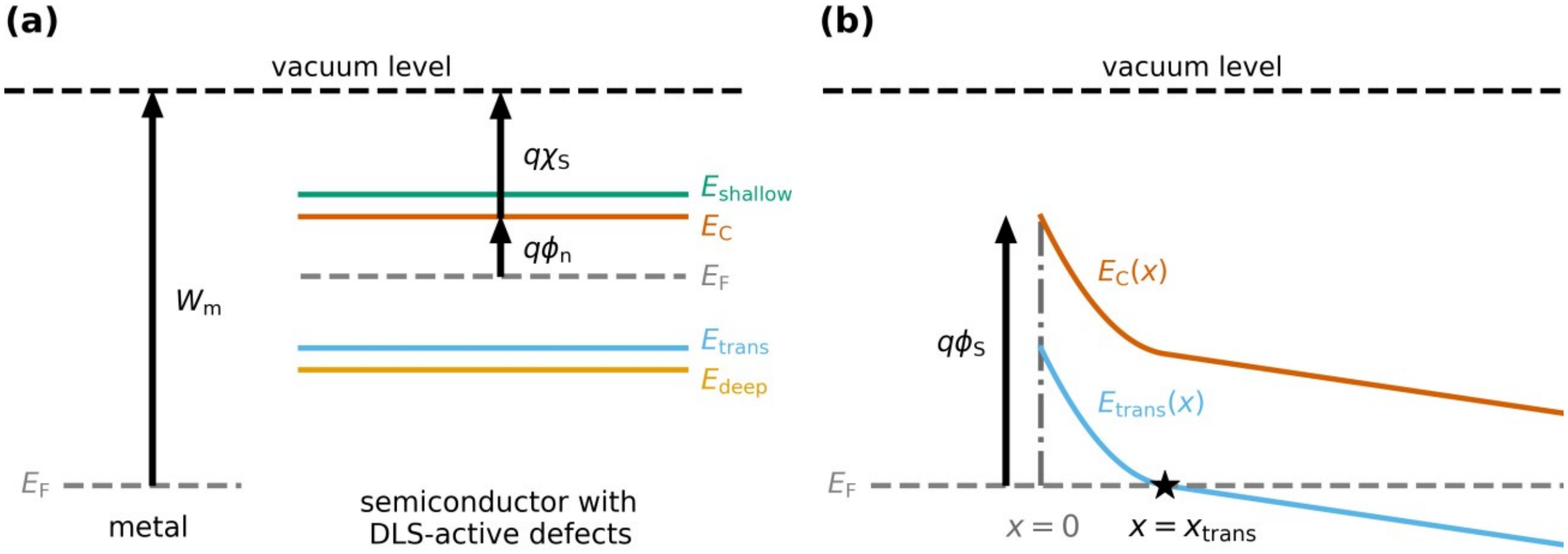

**Figure 2:** Schematic of a DLS-heterojunction. (a) Illustration of energy levels of the materials used: an *n*-type semiconductor with DLS-active defects, and a metal or highly-doped semiconductor. Terms are described in the text. (b) Illustration of the energy band diagram after junction formation. The black arrow indicates the height of the Schottky barrier $q\phi_S$. The gray dash-dotted line indicates the location of the interface where $x = 0$. The black star represents $x = x_{trans}$, *i.e.*, the intersection of the transition energy $E_{trans}(x)$ and Fermi level $E_F$. For $x < x_{trans}$, $E_{trans}(x) > E_F$ and the shallow donor state is thermodynamically favored; while for $x > x_{trans}$, $E_{trans}(x) < E_F$ and the deep donor state is thermodynamically favored.

$x_{trans}$ defines where $V_{An}^{\times}$ and $(V_{An}^{\bullet\bullet})^*$ are equally likely at equilibrium. Under time-varying external stimulus, the distribution of metastable states may be driven out-of-equilibrium, and $x_{trans}$ may no longer describe the position where $N_{deep}(x)$ and $N_{shallow}(x)$ are equal to $N_{DLS}/2$. We define $x_{DLS}$ as the instantaneous position where the deep and shallow donor state concentrations are the same:

$$N_{deep}(x_{DLS}) = N_{shallow}(x_{DLS}) = \frac{N_{DLS}}{2} \tag{17}$$

At $x < x_{DLS}$, $N_{shallow}(x) > N_{deep}(x)$; while at $x > x_{DLS}$, $N_{shallow}(x) < N_{deep}(x)$. At thermodynamic equilibrium, $x_{DLS} = x_{trans}$.

Under positive bias applied to the metal side of the heterojunction in **Fig. 2b**, the concentration of electrons in the space charge region of the semiconductor will increase. Consequently, around $x_{DLS}$ where $(V_{An}^{\bullet\bullet})^*$ and $V_{An}^{\times}$ coexist, electron recombination will convert $(V_{An}^{\bullet\bullet})^*$ to $V_{An}^{\times}$ as described in Eq. (4). As a result, fewer defects are in the ionized state $(V_{An}^{\bullet\bullet})^*$ and more defects are in the deep neutral state $V_{An}^{\times}$, and $x_{DLS}$ will move closer to $x = 0$ according to Eq. (17). This process requires thermal activation, and it's thus slow to follow the applied bias, leading to hysteretic redistribution of $N_{deep}(x,t)$ and $N_{shallow}(x,t)$. Hence, $N_{deep}(x,t)$, $N_{shallow}(x,t)$ and $x_{DLS}(t)$ are dynamical, history-dependent variables.

To track the evolution of $N_{\text{deep}}(x,t)$ and $N_{\text{shallow}}(x,t)$, the Poisson equation must be solved at each time to compute $E_{\text{C}}(x,t)$ and $E_{\text{V}}(x,t)$, which are used to calculate the configuration transition rates defined in Eq. (12) and Eq. (13), and then update $N_{\text{deep}}(x,t)$ and $N_{\text{shallow}}(x,t)$ at next time step. Solving the Poisson equation at each time is computationally demanding because it requires numerical iteration. For faster computational simulation of the device dynamics, we simplify the profile of space charge density to a step function, as was proposed by Kimerling for $p^{+}n$ junctions that contain deep traps.[28] Kimerling assumed that deep traps contribute to the space charge only within a region of width $y$ from the interface, which bears the same meaning as the $x_{\text{DLS}}$ in Eq. (17). Hence, the space charge density is given by the following expression:

$$\rho(x,t) = \begin{cases} 2qN_{\text{DLS}}, & x < x_{\text{DLS}} \\ 0, & x > x_{\text{DLS}} \end{cases} \tag{18}$$

For the cases simulated here, DLS-active defects are chosen to be the dominant type of defects in the materials discussed, *i.e.*, $N_{\text{DLS}}$ is much larger than the shallow (non-DLS) donor concentration $N_{\text{d}}$ (**Table 1**), and the contribution from regular shallow donors to the space charge profile is negligible. For cases in which both metastable and shallow defects make meaningful contributions to the space charge profile, as in CIGS thin films, Decock *et al.* proposed to use a staircase space charge density profile (this multi-step model can be seen as an extension of the one-step model).[14] In **§A2** we explicitly compare the results of this staircase approximate model to direct numerical simulation and find good agreement. Therefore, the assumption of step-function changes in space charge density profile is generally applicable to junctions with deep and shallow levels.

We use Gauss's law to calculate the electric potential profile variation through the junction. The boundary conditions of the conduction band energy $E_{\text{C}}(x,t)$ (referenced to the Fermi energy) on the two sides of the semiconductor are:

$$E_{\text{C}}(x=0,t) = q\phi_{\text{S}}, \qquad E_{\text{C}}(x=L_{\text{s}},t) = q(\phi_{\text{n}} + V_{\text{bias}}) \tag{19}$$

$\varepsilon_{\text{s}}$ and $L_{\text{s}}$ are the dielectric constant and width of the semiconductor, respectively. $V_{\text{bias}}$ is the applied bias at the given time $t$. $E_{\text{C}}(x,t)$ is then given by:

$$E_{\text{C}}(x,t) = \begin{cases} q\phi_{\text{S}} - \dfrac{2N_{\text{DLS}}q^2}{\varepsilon_{\text{s}}}\left[\left(x_{\text{DLS}}(t) + \dfrac{W^2(V_{\text{bias}}) - x_{\text{DLS}}^2(t)}{2L_{\text{s}}}\right)x - \dfrac{x^2}{2}\right], & x < x_{\text{DLS}}(t) \\ \dfrac{N_{\text{DLS}}q^2}{\varepsilon_{\text{s}}}\dfrac{W^2(V_{\text{bias}}) - x_{\text{DLS}}^2(t)}{L_{\text{s}}}(L_{\text{s}} - x) + q(\phi_{\text{n}} + V_{\text{bias}}), & x > x_{\text{DLS}}(t) \end{cases} \tag{20}$$

where $W(V_{\text{bias}})$ is given by:

$$W(V_{\text{bias}}) = \sqrt{\frac{\varepsilon_{\text{s}}}{N_{\text{DLS}}q}[\phi_{\text{S}} - (\phi_{\text{n}} + V_{\text{bias}})]} \tag{21}$$

Taking the parameters of the $MoO_3$/CdS heterojunction (that we list in **Table 1**, and for $MoO_3$ we take its work function to be 6.7 eV), we illustrate the effect of bias on the interface band diagram and metastable defect states in **Fig. 3**. As shown in **Fig. 3a** and **Fig. 3b** are the transition energy level $E_{\mathrm{trans}}(x)$ and the concentration of shallow donor states $N_{\mathrm{shallow}}(x)$ in the CdS layer. $E_{\mathrm{trans}}(x)$ is computed by combining Eq. (11) and Eq. (20) while $N_{\mathrm{shallow}}(x)$ is computed from Eq. (15). In **Fig. 3**: the dashed blue line indicates $E_{\mathrm{trans}}(x)$ or $N_{\mathrm{shallow}}(x)$ at zero bias, while the orange and green solid lines indicate those after forward bias treatment (+0.8 V) and reverse bias treatment (-0.8 V), respectively; in **Fig. 3b**, the vertical dashed lines indicate positions of $x_{\mathrm{DLS}}$. After forward bias treatment, $x_{\mathrm{DLS}}$ moves from 16.1 nm to 9.5 nm. According to Eq. (17), the decrease of $x_{\mathrm{DLS}}$ means fewer defects are in the ionized shallow state, which reduces the effective donor doping concentration and the electric field at the junction; as a result, the interface barrier is thicker than before, switching the junction to high resistance state (HRS). After reverse bias treatment, $x_{\mathrm{DLS}}$ moves from 16.1 nm to 20.7 nm, meaning that more defects are in the ionized state, increasing the effective donor doping concentration and the electric field at the junction; therefore, the interface barrier is thinner than before, switching the junction to low resistance state (LRS). These simulation results are consistent with the bipolar resistance switching that we demonstrated experimentally: DLS devices switch to HRS under forward bias, and switch to LRS under reverse bias.[6] We note that the changes of defect charge states and depletion region upon bias treatment shown in **Fig. 3** are in analogy to those in deep level transient spectroscopy (DLTS) measurements. In DLTS, a forward bias treatment can neutralize defects and widen the depletion region, causing a temporary decrease of the junction capacitance; in this work, the thickening of interface barrier induced by forward bias treatment causes a decrease of the junction conductivity.

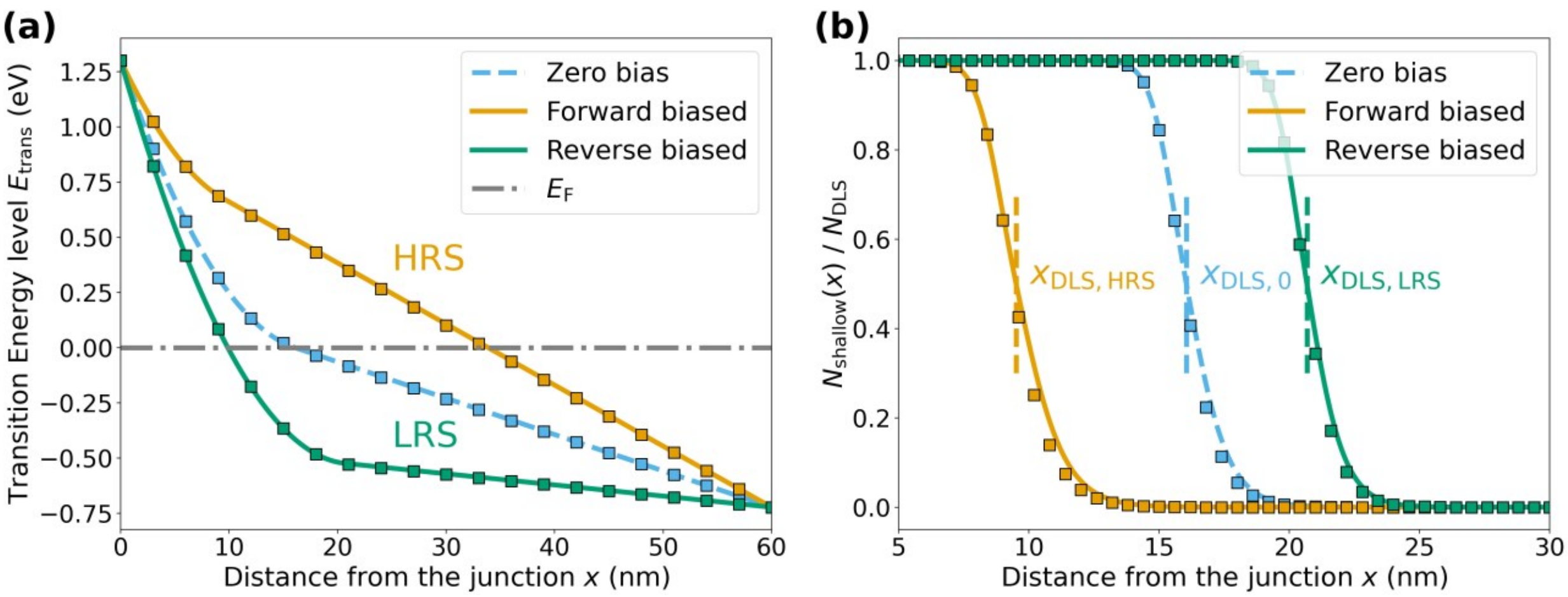

**Figure 3:** Relationship between interface band diagram and metastable defect states. (a) Transition energy level $E_{\mathrm{trans}}(x)$ that is computed by combining Eq. (11) and Eq. (20): without bias treatment (blue dashed line), after forward bias (orange solid line) and reverse bias treatments (green solid line). The square symbols represent the results that we compute using the numerical iteration method developed by Decock *et al.*[14] Energy is referenced to the Fermi level (grey dot-dash line). (b) Distribution of shallow donor states $N_{\mathrm{shallow}}(x)$ that is computed from Eq. (15): without bias treatment (blue dashed line), after forward bias (orange solid line) and reverse bias treatments (green solid line). The vertical dashed lines indicate positions of $x_{\mathrm{DLS}}$. The heterojunction is in HRS and LRS after forward and reverse bias treatments, respectively, and the corresponding positions of $x_{\mathrm{DLS}}$ are labeled as $x_{\mathrm{DLS,HRS}}$ and $x_{\mathrm{DLS,LRS}}$, respectively.

To examine the accuracy and efficiency of our model, we use the numerical iteration method developed by Decock *et al.* to compute $E_{\mathrm{trans}}(x)$ and $N_{\mathrm{shallow}}(x)$.[14] The results are shown as square symbols in **Fig. 3** and they are in good agreement with those from our model which are represented by dashed and solid lines. Also, we estimate that our analytical model is at least one million times more efficient than the numerical iteration approach: to compute one band diagram in **Fig. 3a**, it takes less than 60 micro-seconds with Eq. (20), while it takes about 60 seconds to achieve the self-consistency with the numerical iteration method. All our simulations are executed with Python code and NumPy package.

Using the interface band diagram that we compute from Eq. (20), we estimate that after forward bias treatment the junction resistivity is 783 times that under zero bias (conductivity modeling described in **§A3**). At the doping concentration $N_{\mathrm{DLS}} = 2 \times 10^{18}\ \mathrm{cm}^{-3}$, as in **Table 1**, tunneling is the dominant transport mechanism, and it is exponentially dependent on both of the height and width of the interface barrier. In experiments, we found that the $MoO_3$/CdS heterojunction resistance could increase by up to 200 times after a forward bias treatment of +0.8 V at room temperature.[6] When the concentration of DLS-active defects is reduced $N_{\mathrm{DLS}}$ to $1.5 \times 10^{18}\ \mathrm{cm}^{-3}$, we calculate the resistivity ratio of the forward-biased to the zero-biased junction to be 187, which is close to the experimental value.

As shown in **Fig. 3b**, in the HRS more metastable defects are in the neutral deep state (more electrons are trapped at the interface) while in the LRS more are in the ionized shallow donor configuration (fewer electrons are trapped at the interface). The physical insight is the same as presented by Simmons and Verderber: trapped charge distorts the interface band structure, and as a consequence changes the electric field and tunneling currents passing through the interface.[1]

Eq. (20) and **Fig. 3** indicates that $x_{\mathrm{DLS}}$ solely determines the interface band diagram as well as the resistivity. Therefore, $x_{\mathrm{DLS}}$ is a state variable of the DLS device and the device dynamics are represented by the movement of $x_{\mathrm{DLS}}(t)$. Its velocity, $v_{\mathrm{DLS}}(t)$, can be determined by the space and time variation of $N_{\mathrm{shallow}}(x,t)$ at $x_{\mathrm{DLS}}(t)$:

$$v_{\mathrm{DLS}}(t) = \frac{\partial x_{\mathrm{DLS}}(t)}{\partial t} = -\frac{\dfrac{\partial N_{\mathrm{shallow}}(x_{\mathrm{DLS}},t)}{\partial t}}{\dfrac{\partial N_{\mathrm{shallow}}(x_{\mathrm{DLS}},t)}{\partial x}} \tag{22}$$

Combining Eqs. (9)-(12) and (17), and estimating $N_{\mathrm{C}} = N_{\mathrm{V}}$, we have the numerator of Eq. (22) given by the following expression:

$$\frac{\partial N_{\mathrm{shallow}}(x_{\mathrm{DLS}},t)}{\partial t} = \frac{N_{\mathrm{DLS}}}{v_{\mathrm{ph}}}(V_{\mathrm{th}} N_{\mathrm{C}} \sigma)^2 \exp\left(-\frac{E_{\mathrm{g}} + (\Delta E_{\mathrm{2EC}} + \Delta E_{\mathrm{2HC}})/2}{kT}\right) \sinh\left(\frac{2(E_{\mathrm{trans}}(x_{\mathrm{DLS}},t) - E_{\mathrm{F}})}{kT}\right) \tag{23}$$

Given the boundary conditions of $E_{\mathrm{C}}(x,t)$ in Eq. (19), we have $E_{\mathrm{F}} = qV_{\mathrm{bias}}$, which indicates that the numerator of Eq. (22) can be increased exponentially under applied bias.

We rewrite the denominator of Eq. (22) as:

$$\frac{\partial N_{\text{shallow}}(x_{\text{DLS}}, t)}{\partial x} = -\frac{N_{\text{DLS}}}{L_{\text{C}}} \tag{24}$$

$L_{\text{C}}$ represents the characteristic length of the truncation of the shallow donor distribution $N_{\text{shallow}}(x)$ around $x_{\text{DLS}}$. For the $N_{\text{shallow}}(x)$ shown in **Fig. 3b**, we calculate $L_{\text{C}}$ to be 3.1 nm for the junction under zero bias, and 3.6 nm and 2.8 nm for that after forward and reverse bias treatment, respectively. Since $L_{\text{C}}$ only changes slightly after bias treatment, we take it to be constant and thus Eq. (24) is constant. Combining Eqs. (22)-(24) yields:

$$v_{\text{DLS}}(t) = v_0 \exp\left(-\frac{E_{\text{g}} + \frac{\Delta E_{2\text{EC}} + \Delta E_{2\text{HC}}}{2}}{kT}\right) \sinh\left(\frac{2(E_{\text{trans}}(x_{\text{DLS}}, t) - qV_{\text{bias}})}{kT}\right) \tag{25}$$

where $v_0$ is a constant velocity, that is given by:

$$v_0 = \frac{L_{\text{C}}}{\nu_{\text{ph}}} (V_{\text{th}} N_{\text{C}} \sigma)^2 \tag{26}$$

In deriving Eq. (25), we have simplified the complex process of defect ionization and electron recapture in DLS devices into the motion of $x_{\text{DLS}}$, while preserving the essential switching physics. The expression of the velocity of $x_{\text{DLS}}$ has a similar form to that describing the growth velocity of conductive filaments in metal oxide resistive switching devices: both have a hyperbolic-sinusoidal dependence on the applied bias, which means that in both cases the resistive switching is bipolar and the switching speed can be increased exponentially with the drive voltage amplitude.[39] However, the switching physics is different: in the case of DLS-heterojunction devices, the exponential dependence stems from electron or hole injection whose densities are exponentially dependent upon applied bias, while in the case of metal oxide resistive switching devices, the exponential dependence originates from electric field driven ion hopping. Also, in Eq. (25) all of the parameters are constructed from point defect physics, and there is no mass transport.

### 3. Dynamic behavior of DLS-heterojunction devices upon voltage stimulus

As an example, in **Fig. 4a**, we show the response of $x_{\text{DLS}}$ (red solid line) upon a sinusoidal driving voltage (blue solid line) at a frequency of 1 MHz and an amplitude of 1.8 V. We also show the steady-state (*i.e.*, equilibrium) position of $x_{\text{DLS}}$ upon the driving voltage with the gray dashed line, and we see that the device operates in out-of-equilibrium states of $x_{\text{DLS}}$. As illustrated in **Fig. 4b**, $x_{\text{DLS}}$ is hysteretic with respect to the driving voltage, and it changes in the interval of 13.8 - 17.4 nm. When the device is at HRS, $x_{\text{DLS}} = x_{\text{DLS,HRS}} = 13.8$ nm, and when the device is at LRS, $x_{\text{DLS}} = x_{\text{DLS,LRS}} = 17.4$ nm. Under this sinusoidal driving voltage, we estimate the resistivity ratio of HRS to LRS to be 50.

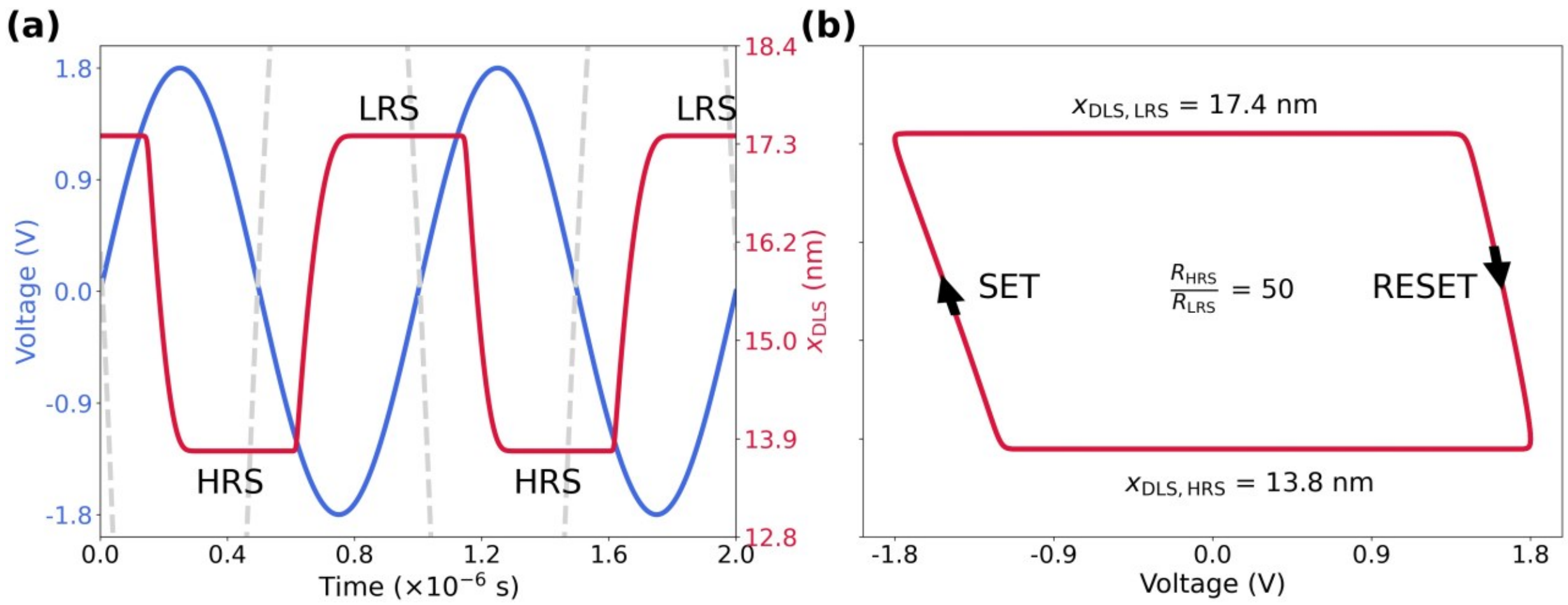

**Figure 4:** Resistive switching and hysteresis behavior of a DLS-heterojunction device under a sinusoidal driving voltage whose frequency is 1 MHz. In (a) we plot the voltage stimulus (blue solid line) and the corresponding change of $x_{\mathrm{DLS}}$ (red solid line) versus time. We also plot the equilibrium position of $x_{\mathrm{DLS}}$ (gray dashed line) versus time. The device operates in out-of-equilibrium states of $x_{\mathrm{DLS}}$. In (b) we plot $x_{\mathrm{DLS}}$ with respect to the voltage stimulus. We label the processes of SET (from HRS to LRS) and RESET (from LRS to HRS) with black arrows.

Since the hysteretic resistive changes are based on non-equilibrium charge states of metastable point defects, DLS devices are predicted to be volatile and not suitable for nonvolatile memories.[15] At lower driving frequency, we expect reduced resistive switching because the metastable defects have more time to approach their thermodynamic equilibrium states. For sufficiently low frequency (smaller than the order of magnitude of the inverse of the device retention time), the hysteretic loop shown in **Fig. 4** will collapses into a line, *i.e.*, the hysteresis vanishes and $x_{\mathrm{DLS}}$ closely follows the applied bias. The room-temperature retention time of the HRS at $MoO_3$/CdS heterojunctions has been found to be about 90 s.[6] At the HRS, $x_{\mathrm{DLS,HRS}} < x_{\mathrm{DLS,0}}$ (as previously demonstrated in **Fig. 3b**), and the DLS-active defects will gradually recover their thermodynamic equilibrium states, by emptying the occupied electrons through electron emission. Using Eq. (5) and the experimental fact that $\tau_{2\mathrm{EE}} = 90$ s, we estimate $\Delta E_{2\mathrm{EE}} = 0.9$ eV. $\tau_{2\mathrm{EE}}^{-1}$ impacts the device dynamics at low driving frequencies. Adding Eq. (5) into Eq. (10) and Eq. (12), we can study the hysteretic behavior of $x_{\mathrm{DLS}}$ at different frequencies and compute the corresponding resistivity ratio of HRS to LRS. In **Fig. 5** we show the resistivity ratio $R_{\mathrm{HRS}}/R_{\mathrm{LRS}}$ upon sinusoidal driving voltage of different frequencies; insets show hysteretic loops of $x_{\mathrm{DLS}}$. At frequency $f = 10^{-4}$ Hz, which is much lower than the inverse of the retention time, *i.e.*, $1/90\ \mathrm{s}^{-1}$, the hysteretic loop of $x_{\mathrm{DLS}}$ collapses into a line as expected, and $R_{\mathrm{HRS}}/R_{\mathrm{LRS}} = 1$. With increasing frequency, $R_{\mathrm{HRS}}/R_{\mathrm{LRS}}$ increases to a maximum and then decreases. The decrease is because, at very high frequency, defects have insufficient time to switch between configurations. As indicated by Eq. (25), the equation of motion of $x_{\mathrm{DLS}}$, the resistivity ratio is limited by the single electron/hole emission rates, the configuration transition barriers, the driving voltage amplitude, as well as other material properties. Using higher voltage amplitude will increase the injected carrier concentration and thus increase the resistivity ratio $R_{\mathrm{HRS}}/R_{\mathrm{LRS}}$.

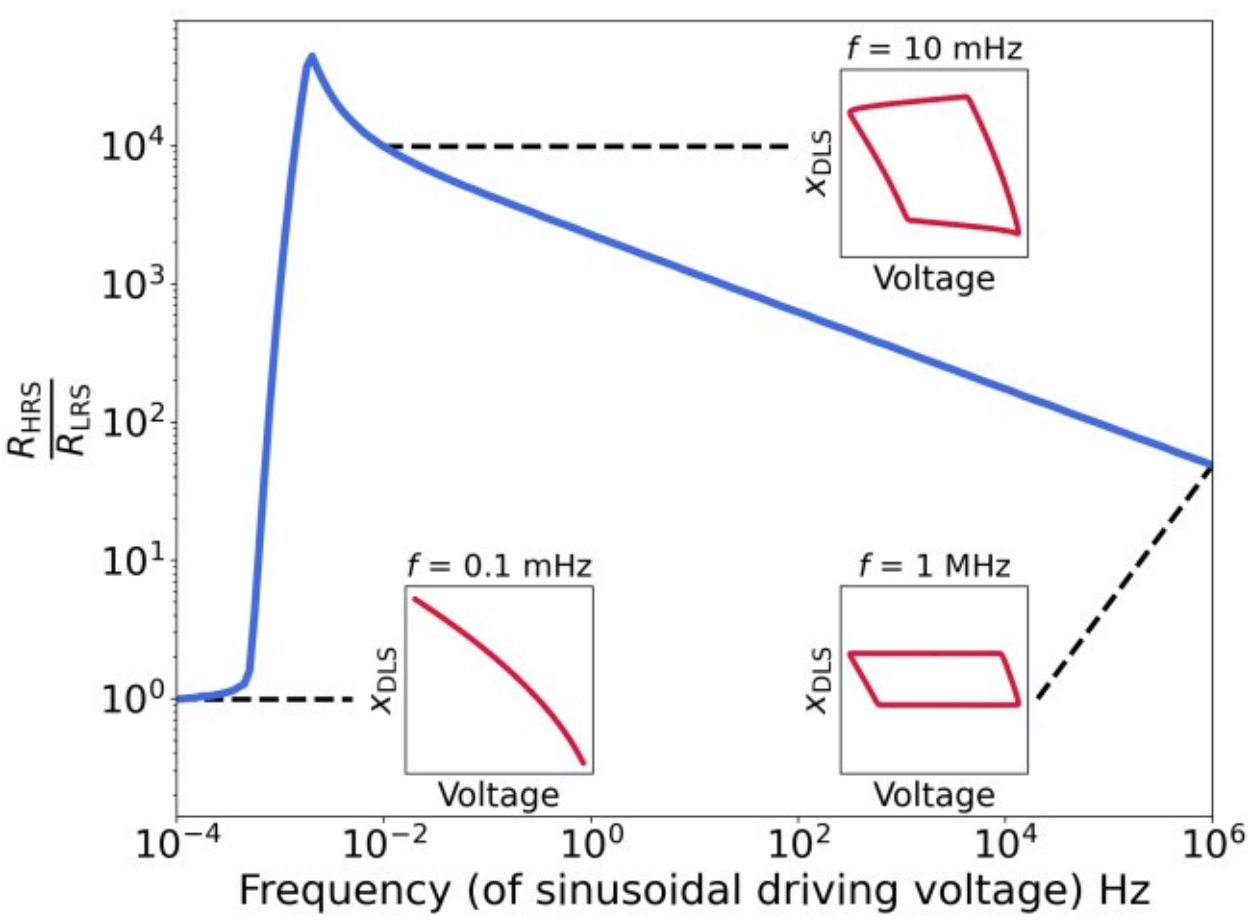

**Figure 5:** The resistivity ratio of HRS to LRS upon sinusoidal driving voltage of frequencies from 0.1 mHz to 1 MHz. In the three insets we plot the hysteretic loop of $x_{\mathrm{DLS}}$ (*i.e.*, the change of $x_{\mathrm{DLS}}$ with respect to the voltage stimulus) at three different frequencies: 0.1 mHz, 10 mHz and 1 MHz.

Using Eq. (25), we can study the switching time between HRS and LRS of DLS-devices under voltage pulses, which are typically used to program resistive states in RRAM devices, and are relevant to neuromorphic computing strategies. To understand the switching process intuitively, we derive analytical expressions of the switching time. Eq. (25) is analytically solvable if we expand $E_{\mathrm{trans}}(x_{\mathrm{DLS}}, t)$ linearly with respect to $x_{\mathrm{DLS}}$, yielding:

$$E_{\mathrm{trans}}(x_{\mathrm{DLS}}, t) = \left(\frac{\partial E_{\mathrm{trans}}}{\partial x_{\mathrm{DLS}}}\right)_0 \left(x_{\mathrm{DLS}} - x_{\mathrm{DLS},0}\right) + \frac{x_{\mathrm{DLS}}}{L_{\mathrm{s}}} qV_{\mathrm{bias}} \tag{27}$$

where $x_{\mathrm{DLS},0}$ is the equilibrium position of $x_{\mathrm{DLS}}$ at zero bias (as previously demonstrated in **Fig. 3b**). Eq. (27) approximates $E_{\mathrm{trans}}(x_{\mathrm{DLS}}, t)$ fairly well in the range of $x_{\mathrm{DLS}}$ and q$V_{\mathrm{bias}}$ that we are interested in (**§A3**).

Under reverse voltage pulse $V_{\mathrm{bias}} < 0$, the device changes from HRS to LRS (the SET operation): $x_{\mathrm{DLS}}$ moves from $x_{\mathrm{DLS,HRS}}$ to $x_{\mathrm{DLS,LRS}}$. Combining Eq. (25) and Eq. (27), we solve for the switching time $t_{\mathrm{SET}}$:

$$t_{\mathrm{SET}} = \frac{1}{-\frac{1}{kT}\left(\left(\frac{\partial E_{\mathrm{trans}}}{\partial x_{\mathrm{DLS}}}\right)_0 + \frac{qV_{\mathrm{bias}}}{L_{\mathrm{s}}}\right) v_0} \exp\left(\frac{E_{\mathrm{g}} + \frac{\Delta E_{2\mathrm{EC}} + \Delta E_{2\mathrm{HC}}}{2}}{kT}\right)$$
$$\exp\left(-\frac{2}{kT}\left[q|V_{\mathrm{bias}}|\left(1 - \frac{x_{\mathrm{DLS,LRS}}}{L_{\mathrm{s}}}\right) + \left(\frac{\partial E_{\mathrm{trans}}}{\partial x_{\mathrm{DLS}}}\right)_0 \left(x_{\mathrm{DLS,LRS}} - x_{\mathrm{DLS},0}\right)\right]\right) \tag{28}$$

Under forward voltage pulse $V_{\mathrm{bias}} > 0$, the device changes from LRS to HRS (the RESET operation): $x_{\mathrm{DLS}}$ moves from $x_{\mathrm{DLS,LRS}}$ to $x_{\mathrm{DLS,HRS}}$. Combining Eq. (25) and Eq. (27), we solve for the switching time $t_{\mathrm{RESET}}$:

$$t_{\text{RESET}} = \frac{1}{-\frac{1}{kT}\left(\left(\frac{\partial E_{\text{trans}}}{\partial x_{\text{DLS}}}\right)_0 + \frac{qV_{\text{bias}}}{L_{\text{s}}}\right) v_0} \exp\left(\frac{E_{\text{g}} + \frac{\Delta E_{\text{2EC}} + \Delta E_{\text{2HC}}}{2}}{kT}\right)$$
$$\exp\left(-\frac{2}{kT}\left[q|V_{\text{bias}}|\left(1 - \frac{x_{\text{DLS,HRS}}}{L_{\text{s}}}\right) + \left(\frac{\partial E_{\text{trans}}}{\partial x_{\text{DLS}}}\right)_0 \left(x_{\text{DLS,0}} - x_{\text{DLS,HRS}}\right)\right]\right) \tag{29}$$

We see in Eq. (28) and Eq. (29) that the SET/RESET switching times can be exponentially reduced by increasing the pulse amplitude $|V_{\text{bias}}|$. So, using voltage pulses of larger amplitude is preferred to achieve faster switching. We note that $t_{\text{SET}}$ is only dependent on $x_{\text{DLS,LRS}}$ and independent of $x_{\text{DLS,HRS}}$; vice versa, $t_{\text{SET}}$ is solely dependent on $x_{\text{DLS,HRS}}$, regardless of the position of $x_{\text{DLS,LRS}}$.

Eq. (28) and Eq. (29) enable predictive device design and selection of operating parameters. Provided that we are required to design a device with a resistance ratio of 50. From **Fig. 4** we learn that using a combination of $x_{\text{DLS,HRS}} = 13.8$ nm and $x_{\text{DLS,LRS}} = 17.4$ nm can achieve a resistance ratio of 50. Then, to switch the $x_{\text{DLS}}$ between 13.8 nm and 17.4 nm, we need to compute the time widths ($t_{\text{SET}}$ and $t_{\text{RESET}}$) and amplitude ($|V_{\text{bias}}|$) of programming pulses.

In **Fig. 6a** we compute and plot $t_{\text{SET}}$ and $t_{\text{RESET}}$ as a function of $|V_{\text{bias}}|$, using Eq. (28) and Eq. (29); the pulse time widths and amplitude should be chosen accordingly. For instance, we use a pulse amplitude of $|V_{\text{bias}}| = 1.8$ V; then the pulse time widths need to be $t_{\text{SET}} = 50.0$ ns and $t_{\text{RESET}} = 47.2$ ns, respectively. In **Fig. 6b** we demonstrate the dynamic evolution of $x_{\text{DLS}}$ under a train of such SET and RESET voltage pulses, and the $x_{\text{DLS}}$ starts from the equilibrium position $x_{\text{DLS,0}}$, and switches between $x_{\text{DLS,LRS}}$ and $x_{\text{DLS,HRS}}$ as designed. Here, the evolution of $x_{\text{DLS}}$ is computed with Eq. (25).

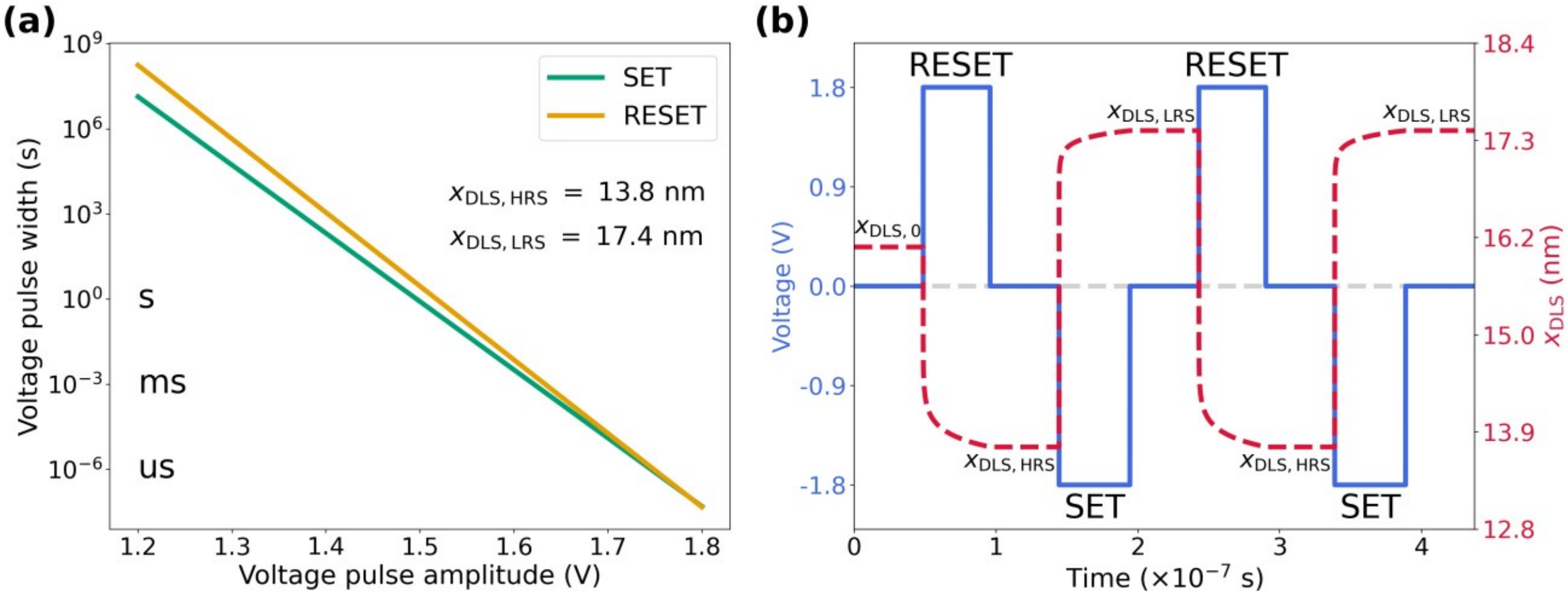

Figure 6: SET/RESET of a DLS-heterojunction device with a resistance ratio of 50, triggered by voltage pulses, and $x_{\text{DLS}}$ of HRS and LRS are set to be 13.8 nm and 17.4 nm, respectively, *i.e.*, $x_{\text{DLS,HRS}} = 13.8$ nm and $x_{\text{DLS,LRS}} = 17.4$ nm. (a) Time widths of voltage pulse required to trigger the SET (green solid line) and RESET (orange solid line) versus amplitudes of voltage pulse. (b) A train of SET and RESET voltage pulses (blue solid line) that switch the device between HRS and LRS. The voltage pulses have an amplitude of $|V_{\mathrm{bias}}| = 1.8$ V and time widths of $t_{\mathrm{SET}} = 50.0$ ns and $t_{\mathrm{RESET}} = 47.2$ ns, respectively. The grey dashed line indicates the position where the voltage is zero. We compute and plot the dynamic evolution of $x_{\mathrm{DLS}}$ (red dashed line) using Eq. (25). $x_{\mathrm{DLS}}$ starts from $x_{\mathrm{DLS},0} = 16.1$ nm, which is the equilibrium position of $x_{\mathrm{DLS}}$ at zero voltage, and it then switches between $x_{\mathrm{DLS,HRS}}$ and $x_{\mathrm{DLS,LRS}}$ as designed.

## 4. Discussion and conclusion

In this work, we use the statistics of metastable point defects and an approximation of step-changes in charge density to develop a physics-based model of resistive switching dynamics in DLS-based heterojunction devices, as we previously demonstrated experimentally.[6] The analytical nature of our model allows for highly efficient device simulations, compared to iterative methods, and we validate our model approximations by reference to the numerical iteration. We explore how devices respond to voltage pulses and variable-frequency drive.

DLS devices are based on charge trapping/detrapping at immobile point defects, and thus the relevant resistive switching mechanism is purely electronic and inherently volatile. Therefore, in DLS devices the programmed resistive state will relax into a thermodynamically stable state after removing the programming signal, but the relaxation may offer desirable dynamics for the emulation of biological neurons and synapses.[24] A notable feature of DLS devices is that they don't suffer from stochastic ion migration events which occur in devices based on mass transport, such as conductive filament (CF) based RRAM.[40] As a result, DLS devices are more amenable to physics-based design and are likely less affected by cycle-to-cycle variation and offer long-term stability.

Predictive modeling of these volatile resistive switching devices will help in designing computing circuits that leverage their dynamics. For future work, a computational database of DLS-active defects is required. The database should include defect parameters which are necessary to guide the design of DLS devices, and these parameters can be computed by DFT calculations and calibrated by experiments. Such a computational database of defects, combined with the physics-based model presented here, suggests a future in which circuits employing DLS devices could be optimized through an inverse design process: first by identifying the circuit dynamics that best suit the proposed algorithm, then resolving dynamics to the level of individual devices, and then achieving the desired device dynamics by materials selection. Volatile DLS devices with designer dynamics could be useful for circuits of coupled, hysteretic, and nonlinear oscillators, such as concepts in physical reservoir computing.[41] Our model is also useful in other junction devices containing metastable defects, such as solar cells, in which metastable defects have direct impact on device performance.[42]

## Acknowledgments

This work was supported in part by the Office of Naval Research MURI through Grant No. N00014-17-1-2661, and in part by the Advanced Concepts Committee of MIT Lincoln Laboratory.

## Appendices

### A1. Equilibrium distributions of different charge states and configuration transition rates for the oxygen vacancy in ZnO

The singly-ionized state $\mathrm{V}_{\mathrm{An}}^{\bullet}$ is thermodynamically unstable, *i.e.*, the energy of the singly-ionized defect $\mathrm{V}_{\mathrm{An}}^{\bullet}$ is always higher than the neutral state $\mathrm{V}_{\mathrm{An}}^{\times}$ and the metastable state $(\mathrm{V}_{\mathrm{An}}^{\bullet\bullet})^{*}$, as shown in **Fig. A1a**.[26,27] According to the formation energies, we compute the equilibrium concentrations of different charge states of the oxygen vacancy as a function of Fermi energy, as shown in **Fig. A1b**. The dominant charge state is either $\mathrm{V}_{\mathrm{An}}^{\times}$ or $(\mathrm{V}_{\mathrm{An}}^{\bullet\bullet})^{*}$, and the concentration of the singly-ionized state $\mathrm{V}_{\mathrm{An}}^{\bullet}$ is at least 10 orders of magnitude smaller than the dominant charge state. Therefore, the transitions $\mathbf{V}_{\mathbf{An}}^{\times} \rightarrow \mathbf{V}_{\mathbf{An}}^{\bullet} + \boldsymbol{e}$ and $(\mathbf{V}_{\mathbf{An}}^{\bullet\bullet})^{*} + \boldsymbol{e} \rightarrow \mathbf{V}_{\mathbf{An}}^{\bullet}$ are experimentally inaccessible, and we only consider and model the transition rates between neutral state $\mathrm{V}_{\mathrm{An}}^{\times}$ and doubly-ionized metastable state $(\mathrm{V}_{\mathrm{An}}^{\bullet\bullet})^{*}$ in this work.

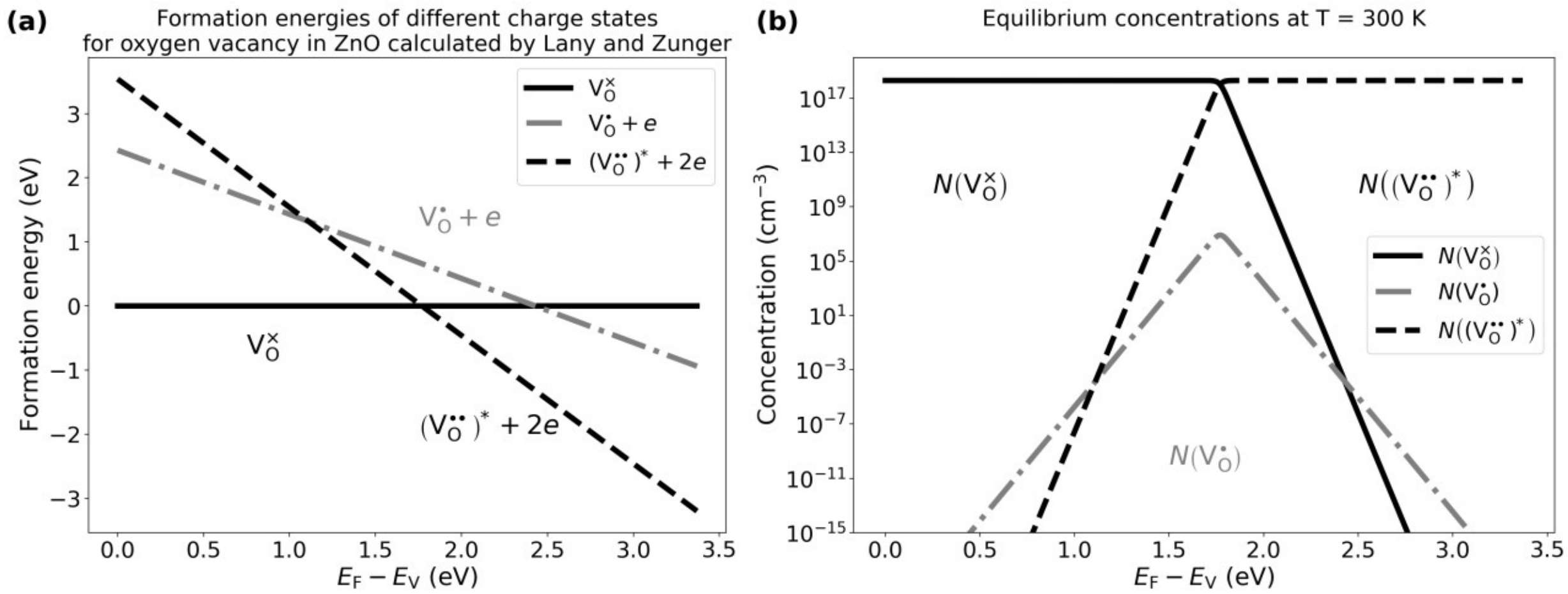

**Figure A1:** (a) Formation energies and (b) equilibrium concentrations of charge states $V_O^\times$, $V_O^\bullet$ and $(V_O^{\bullet\bullet})^*$ for the oxygen vacancy in ZnO as a function of Fermi energy $E_F$ at temperature T=300 K. $E_V$ is the valence band energy level. Formation energies were calculated by Lany and Zunger.[27] Here, the total concentration of oxygen vacancies is $2 \times 10^{18}$ cm$^{-3}$. We use black solid line, black dashed line and grey dash-dotted line to represent the formation energy or equilibrium concentration of charge state $V_O^\times$, $V_O^\bullet$ and $(V_O^{\bullet\bullet})^*$, respectively.

We plot the configuration transition rates $\tau_{2HC}^{-1}$ (green solid line, defined in Eq. (8)), $\tau_{1HC}^{-1}$ (grey solid line, defined in Eq. (7)), $\tau_{2EE}^{-1}$ (blue solid line, defined in Eq. (5)) and $\tau_{2EC}^{-1}$ (orange solid line, defined in Eq. (6)) as a function of Fermi energy $E_F$ at temperature T=300 K and T=1000 K in **Fig. A2a** and **Fig. A2b**, respectively. $\tau_{2HC}^{-1}$, $\tau_{1HC}^{-1}$ and $\tau_{2EE}^{-1}$ can contribute to the deep-to-shallow transition rate $\tau_{D\to S}^{-1}$. As shown in **Fig. A2**, at a high concentration of injected holes, $\tau_{2HC}^{-1}$ is the dominant contribution. Also, the configuration transition rates need to be comparable to the frequency of driving voltage (*i.e.*, 1 MHz $= 10^6$ s$^{-1}$, which is indicated by a black dashed line in **Fig. A2**) to have an impact on the device dynamics. At T=300 K, $\tau_{1HC}^{-1}$ is way lower than $10^6$ s$^{-1}$.

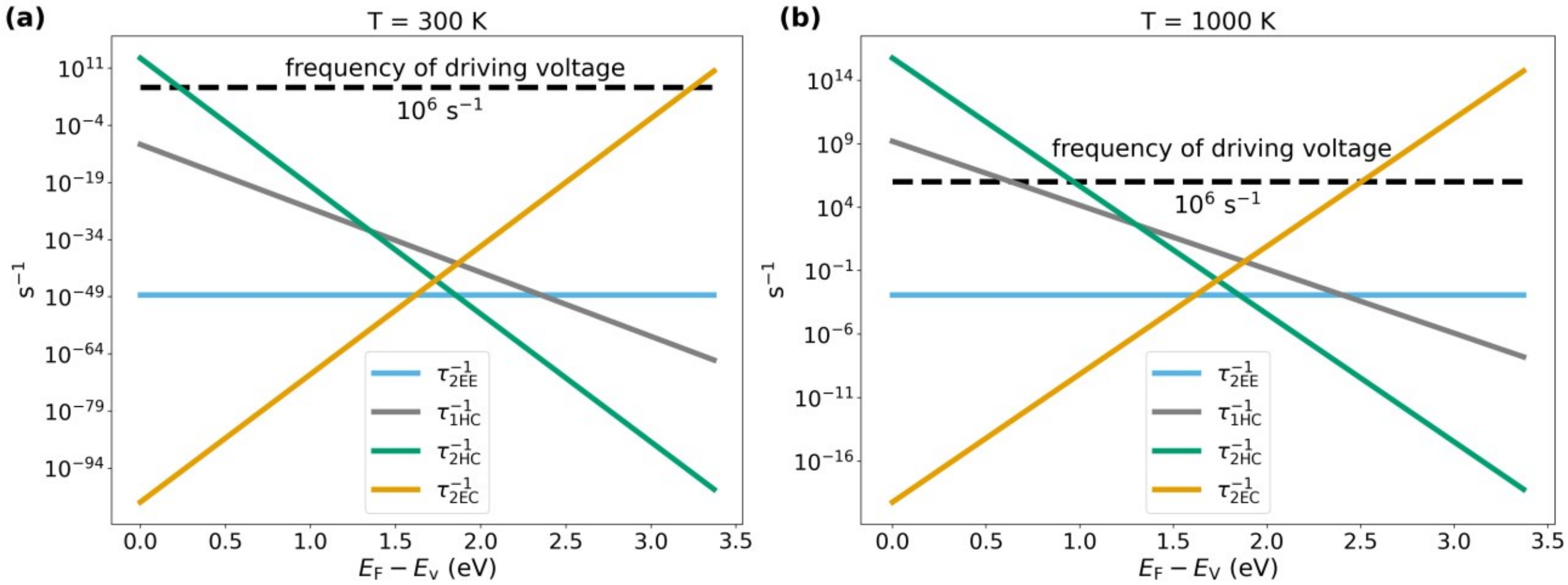

**Figure A2:** Configuration transition rates $\tau_{2HC}^{-1}$ (green solid line), $\tau_{1HC}^{-1}$ (grey solid line), $\tau_{2EE}^{-1}$ (blue solid line) and $\tau_{2EC}^{-1}$ (orange solid line) as a function of Fermi energy $E_F$ at temperature (a) T=300 K and (b) T=1000 K, respectively. $E_V$ is the valence band

energy level. We use a black dashed line to indicate the frequency of driving voltage that we use in device simulations, *i.e.*, 1 MHz $= 10^6$ $s^{-1}$.

**A2. Calculation of apparent defect density profiles using staircase space charge density profile**

We reexamine the calculation of apparent defect density profiles using staircase charge density profile, as proposed by Kimerling and Decock *et al.*[14,28] Decock *et al.* proposed to use this approximation to explain the measurement results of apparent doping density profiles in CIGS solar cells (Eqs. (13), (16) and (17) in ref. [14]), but they did not directly compare the modeling results of this approximate model to their experimental data and to iterative numerical simulations.

The DLS-active defects in CIGS solar cells are $V_{Se} - V_{Cu}$ complexes. We optimize the following four parameters: (1) the built-in barrier of the CdS/CIGS heterojunction, (2) the shallow doping density $N_A$, (3) the concentration of DLS-active defects $N_{DLS}$ (the transition energy level is $E_{trans} - E_V = 0.19$ eV), and (4) the density of the additional acceptor defect $N_t$ (the energy level is $E_t - E_V = 0.33$ eV). The optimized parameters are listed in **Table A1**, together with all the other parameters that are used for the calculation of apparent defect density profiles.

| Materials | ZnO/CdS/CIGS |
|---|---|
| Temperature (K) | 200 |
| Thickness of the CIGS layer (um) | 1.0 |
| Hole effective mass of the CIGS | $0.087m_e$ |
| Built-in barrier of the CdS/CIGS heterojunction (V) | 0.5 |
| $N_A$ ($cm^{-3}$) | $0.4 \times 10^{15}$ |
| DLS-active defects | $V_{Se} - V_{Cu}$ complexes |
| $N_{DLS}$ ($cm^{-3}$) | $0.8 \times 10^{15}$ |
| $E_{trans} - E_V$ (eV) | 0.19 |
| $N_t$ (cm/s) | $1 \times 10^{16}$ |
| $E_t - E_V$ (eV) | 0.33 |

**Table A1:** Overview of the material parameters of CIGS solar cells that we use for the calculation of apparent defect density profiles.

We present in **Fig. A3** the calculated apparent doping density profiles for the CdS/CIGS heterojunction without bias treatment (blue solid line) and after a reverse bias treatment of -2 V (red solid line). The square and triangle symbols represent the numerical simulation and experimental data from Decock *et al.*, respectively; the blue and red colors represent those with and without reverse bias treatment, respectively. The calculated apparent doping density profiles are in satisfactory agreement with the results from Decock *et al.*

For the heterojunction after reverse bias treatment, there is a sharp increase of the apparent doping density (red solid line) when the depletion width reduces below 0.63 um, creating a local maximum of the apparent defect density. The reason is that, in the heterojunction, the length of the distribution of metastable states, that we define as $x_{DLS}$ in the main text, is 0.63 um. When the depletion width is larger than 0.63 um, $x_{DLS}$ is fixed at 0.63 um and DLS-active defects don't respond to the capacitance-voltage testing. When the depletion width shrinks below 0.63 um,

Decock *et al.* assume that these DLS-active defects start to respond to the capacitance-voltage testing as regular shallow dopants (the applied equation for the apparent defect density changes from Eq. (17) to Eq. (16) in ref. [14]), and as a consequence, the apparent density will jump higher; also, $x_{\text{DLS}}$ is considered equal to the depletion width and instantaneously changes upon external bias. A more accurate model would consider the dynamics of $x_{\text{DLS}}$ upon external bias, such as Eq. (25) in this work.

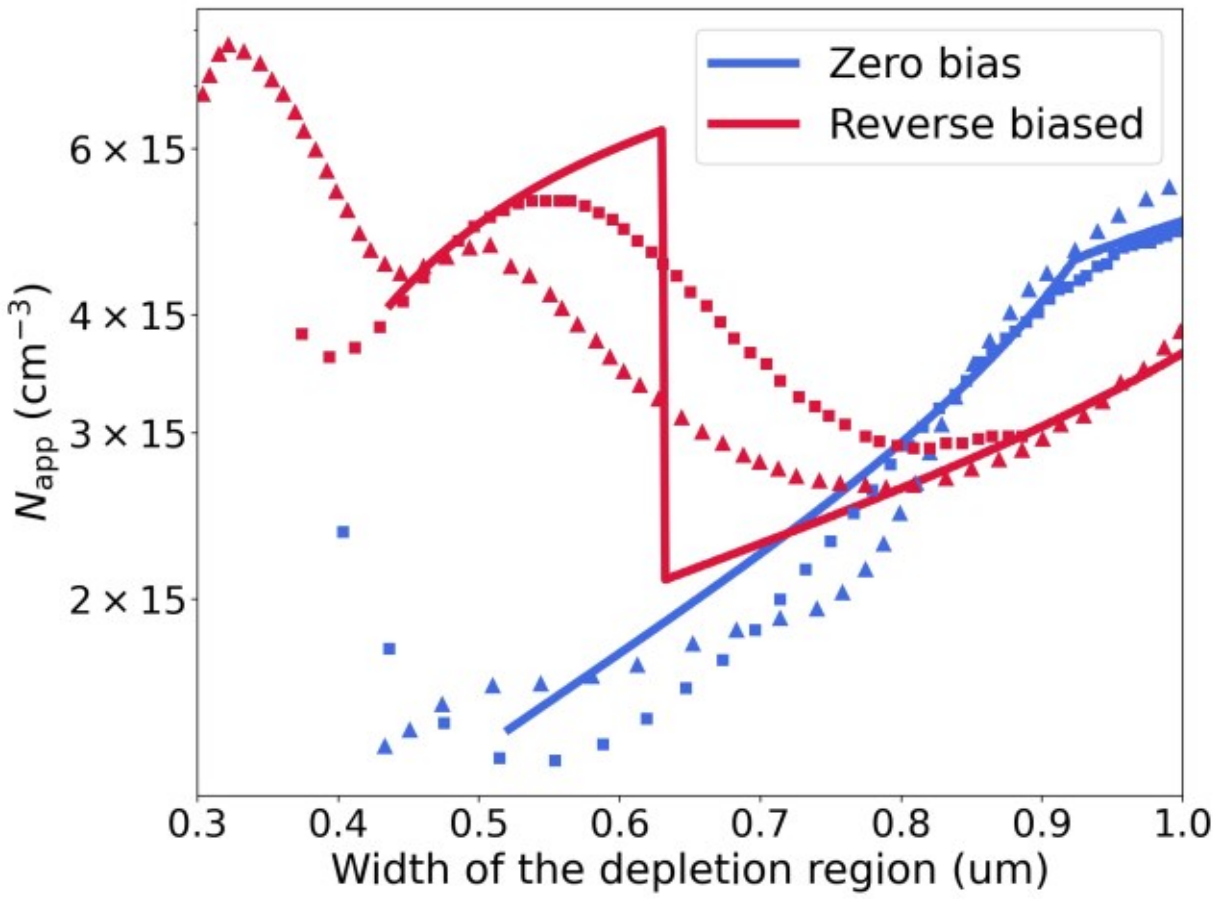

**Figure A3:** Calculated apparent defect density profiles at the CdS/CIGS heterojunction without bias treatment (blue solid line) and after a reverse bias treatment of -2 V (red solid line). Solid lines are the modeling results using the approximation of staircase charge density profiles. The square and triangle symbols represent the numerical simulation results and experimental data from Decock *et al.*, respectively. Blue and red represent with and without reverse bias treatment, respectively.

### A3. Calculation of the junction resistivity

To calculate the junction resistivity, we consider thermionic emission $J_{\text{thermionic}}$, diffusion-drift current $J_{\text{diffusion-drift}}$, and tunneling current $J_{\text{tunneling}}$. The total current density $J$ that flows through the junction is given by:

$$J = J_{\text{thermionic}} + J_{\text{diffusion-drift}} + J_{\text{tunneling}} \tag{A1}$$

From thermionic-emission theory we have:[43]

$$J_{\text{thermionic}} = \frac{qm^*(kT)^2}{2\pi^2\hbar^3} \exp\left(-\frac{q\phi_{\text{S}}}{kT}\right)\left(\exp\left(\frac{qV_{\text{bias}}}{kT}\right) - 1\right) \tag{A2}$$

where $m^*$ is the electron effective mass in the conduction band minimum (CBM), and $q\phi_{\text{S}}$ the Schottky barrier height which has been shown in **Fig. 2** in the main text. We calculate the diffusion-drift current using the following expression:[43]

$$J_{\text{diffusion-drift}} = \mu_{\text{n}} N_{\text{C}} \frac{kT\left(\exp\left(\frac{qV_{\text{bias}}}{kT}\right) - 1\right)}{\int_0^{L_{\text{s}}} \exp\left(\frac{E_{\text{C}}(x)}{kT}\right) dx} \tag{A3}$$

where $\mu_{\mathrm{n}}$ is the electron mobility and $E_{\mathrm{C}}(x)$ is the conduction band level which has been given by Eq. (20). We calculate the tunneling current density using the Tsu Esaki mechanism:[44]

$$J_{\mathrm{tunneling}}=\frac{qm^{*}(kT)^{2}}{2\pi^{2}\hbar^{3}}\int_{0}^{q\phi_{\mathrm{S}}-(q\phi_{\mathrm{n}}+qV_{\mathrm{bias}})}\frac{dE_{\mathrm{x}}}{kT}\ln\left(\frac{1+\exp\left(\frac{-q\phi_{\mathrm{n}}-E_{\mathrm{x}}}{kT}\right)}{1+\exp\left(\frac{-q\phi_{\mathrm{n}}-qV_{\mathrm{bias}}-E_{\mathrm{x}}}{kT}\right)}\right)P(E_{\mathrm{x}}) \quad \text{(A4)}$$

where $E_{\mathrm{x}}=p_{\mathrm{x}}^{2}/2m^{*}$ and $p_{\mathrm{x}}$ is the normal component of the electron momentum that is perpendicular to the heterojunction. $q\phi_{\mathrm{n}}$ has been shown in **Fig. 2** which refers to the energy difference between $E_{\mathrm{C}}$ and the Fermi level $E_{\mathrm{F}}$. $P(E_{\mathrm{x}})$ is ratio of the transmitted to the incident current, and using WKB approximation, $P(E_{\mathrm{x}})$ has been given by Harrison and Stratton:[45,46]

$$P(E_{\mathrm{x}})=\exp\left(-\frac{2\sqrt{2m^{*}}}{\hbar}\int_{x_{1}}^{x_{2}}dx\sqrt{qV_{\mathrm{barrier}}(x)-E_{\mathrm{x}}}\right) \quad \text{(A5)}$$

where $qV_{\mathrm{barrier}}(x)$ is the barrier potential which we compute from the conduction band energy profile $E_{\mathrm{C}}(x)$, and $x_{1}$ and $x_{2}$ are classical turning points what satisfy $qV_{\mathrm{barrier}}(x_{1})-E_{\mathrm{x}}=qV_{\mathrm{barrier}}(x_{2})-E_{\mathrm{x}}=0$.

Combining Eqs. (A1)-(A5), we compute the total current density $J$ within the voltage range [-10 mV, 10 mV], and then we are able to calculate the junction resistance (per area) by linearly fitting the total current density $J$ versus the voltage.

In the calculation of the resistivity ratio, we assume that the Schottky barrier height is constant in both LRS and HRS. In fact, due to the image-force lowering effect, the resistivity ratio $R_{\mathrm{HRS}}/R_{\mathrm{LRS}}$ will be larger than the result calculated here. We denote the image-force-induced lowering is $\Delta\phi_{\mathrm{S}}$. As shown in **Fig. 3**, the interface electric field in the LRS is larger than that in the HRS. Thus, $\Delta\phi_{\mathrm{S,LRS}}>\Delta\phi_{\mathrm{S,HRS}}$ (Eq. (32) in page 148 of ref. [43]), and the Schottky barrier height in the LRS is lower than that in the HRS. This will further increase the resistivity ratio $R_{\mathrm{HRS}}/R_{\mathrm{LRS}}$ compared to the case in which the Schottky barrier is constant in both LRS and HRS.

**A4. Accuracy of the approximate expression of $E_{\mathrm{trans}}(x_{\mathrm{DLS}},t)$**

In Eq. (27), we develop an approximate expression of $E_{\mathrm{trans}}(x_{\mathrm{DLS}},t)$ by expanding it linearly with respect to $x_{\mathrm{DLS}}$ at zero bias. $(\partial E_{\mathrm{trans}}/\partial x_{\mathrm{DLS}})_{0}$ in Eq. (27) is given by the following expression:

$$\left(\frac{\partial E_{\mathrm{trans}}}{\partial x_{\mathrm{DLS}}}\right)_{0}=\frac{N_{\mathrm{DLS}}q^{2}}{\varepsilon_{\mathrm{s}}}\frac{-W^{2}(V_{\mathrm{bias}}=0)-2L_{\mathrm{s}}x_{\mathrm{DLS,0}}+3x_{\mathrm{DLS,0}}^{2}}{L_{\mathrm{s}}} \quad \text{(A6)}$$

Using Eq. (A6), we calculate $(\partial E_{\mathrm{trans}}/\partial x_{\mathrm{DLS}})_{0}=-0.11\ \mathrm{eV/nm}$.

The exact values of $E_{\mathrm{trans}}(x_{\mathrm{DLS}},t)$ can be computed by combining Eq. (11) and Eq. (20). In **Fig. A4**, we compare the approximate values of $E_{\mathrm{trans}}(x_{\mathrm{DLS}},t)$ versus $x_{\mathrm{DLS}}$ and $V_{\mathrm{bias}}$ which are computed from Eq. (27), with the exact values of $E_{\mathrm{trans}}(x_{\mathrm{DLS}},t)$ which are computed by combining Eq. (11) and Eq. (20). In **Fig. A4a** we fix $V_{\mathrm{bias}}=0$, and compute and plot $E_{\text{trans}}(x_{\text{DLS}}, t)$ versus $x_{\text{DLS}}$ within the interval [$x_{\text{DLS,0}}$ - 5 nm, $x_{\text{DLS,0}}$ + 5 nm], *i.e.*, [11.1 nm, 21.1 nm]. In **Fig. A4b** we fix $x_{\text{DLS}} = x_{\text{DLS,0}} = 16.1$ nm, and compute and plot $E_{\text{trans}}(x_{\text{DLS}}, t)$ versus $V_{\text{bias}}$ in the interval of [-1.8 V, 1.8 V]. From **Fig. A4** we conclude that Eq. (27) approximates $E_{\text{trans}}(x_{\text{DLS}}, t)$ fairly well.

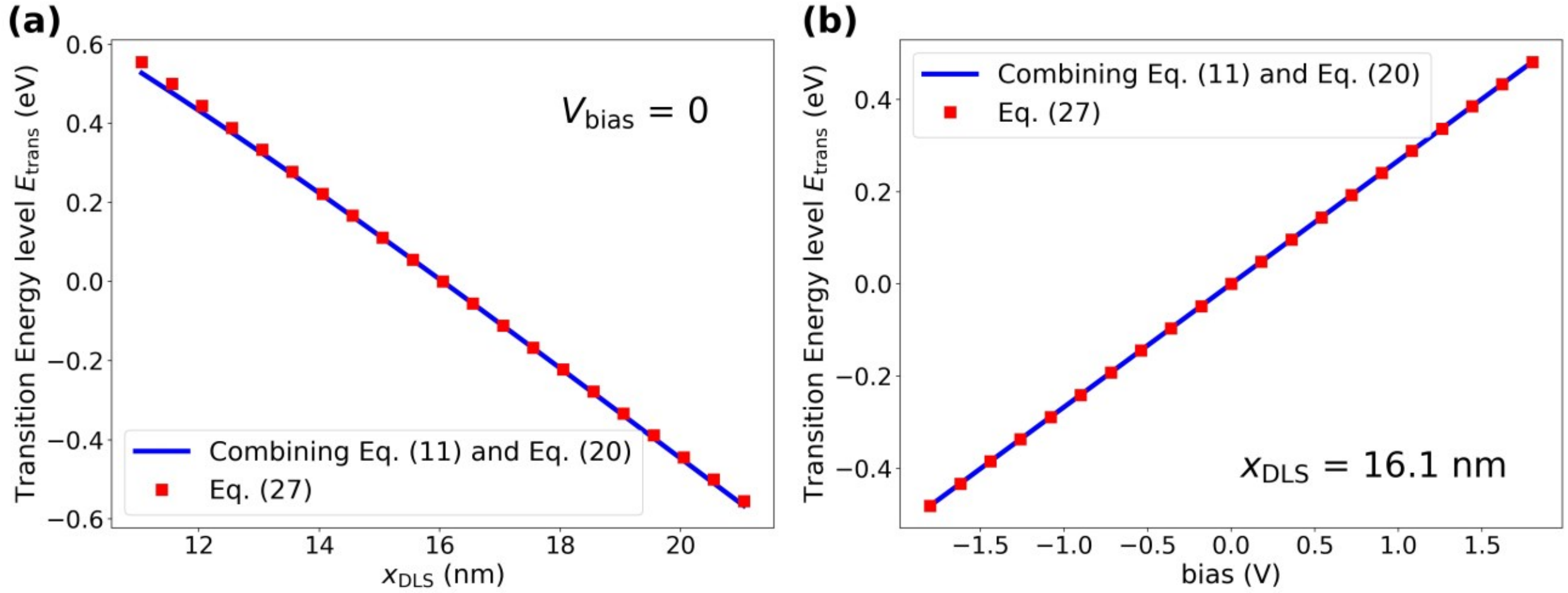

**Figure A4:** Comparison between the approximate values of $E_{\text{trans}}(x_{\text{DLS}}, t)$ (represented by red squares) that are computed from Eq. (27) with the exact values of $E_{\text{trans}}(x_{\text{DLS}}, t)$ (blue solid lines) that are computed by combining Eq. (11) and Eq. (20). In (a) we fix $V_{\text{bias}} = 0$, and compute and plot $E_{\text{trans}}(x_{\text{DLS}}, t)$ versus $x_{\text{DLS}}$ in the interval of [$x_{\text{DLS,0}}$ - 5 nm, $x_{\text{DLS,0}}$ + 5 nm], *i.e.*, [11.1 nm, 21.1 nm]. In (b) we fix $x_{\text{DLS}} = x_{\text{DLS,0}} = 16.1$ nm, and compute and plot $E_{\text{trans}}(x_{\text{DLS}}, t)$ versus $V_{\text{bias}}$ in the interval of [-1.8 V, 1.8 V].